%% file: main.tex
\documentclass{cta-author} 

\usepackage{xcolor}
\usepackage{graphicx}
\usepackage{algorithm}
\usepackage{amsmath}
\usepackage{amssymb}
\usepackage{textcomp}
\usepackage{url}
\usepackage{adjustbox}

\usepackage{xifthen}
\usepackage{multirow}

\usepackage{algorithm,algcompatible}

\usepackage{xcolor}
\definecolor{PlotColorA}{HTML}{204A87}
\definecolor{PlotColorB}{HTML}{677F34}
\definecolor{PlotColorC}{HTML}{C24C5A}
\definecolor{PlotColorD}{HTML}{cee1f3}

\definecolor{ColorOri00}{HTML}{EE55F2}
\definecolor{ColorOri01}{HTML}{FB83CD}
\definecolor{ColorOri02}{HTML}{FAB0A4}
\definecolor{ColorOri03}{HTML}{FED472}
\definecolor{ColorOri04}{HTML}{EFEC36}
\definecolor{ColorOri05}{HTML}{C8E816}
\definecolor{ColorOri06}{HTML}{9AD20F}
\definecolor{ColorOri07}{HTML}{5EC30D}
\definecolor{ColorOri08}{HTML}{36AE26}
\definecolor{ColorOri09}{HTML}{3E8F62}
\definecolor{ColorOri10}{HTML}{41759B}
\definecolor{ColorOri11}{HTML}{274FC5}
\definecolor{ColorOri12}{HTML}{2E24E9}
\definecolor{ColorOri13}{HTML}{6021F7}
\definecolor{ColorOri14}{HTML}{9039F8}
\definecolor{ColorOri15}{HTML}{C043FD}

\definecolor{ColorStrengthTop}{HTML}{007991}
\definecolor{ColorStrengthBottom}{HTML}{f8f8f8}
\definecolor{ColorCoherenceTop}{HTML}{dc6200}
\definecolor{ColorCoherenceBottom}{HTML}{f8f8f8}

\makeatletter
\def\thickhline{%
  \noalign{\ifnum0=`}\fi\hrule \@height \thickarrayrulewidth \futurelet
   \reserved@a\@xthickhline}
\def\@xthickhline{\ifx\reserved@a\thickhline
               \vskip\doublerulesep
               \vskip-\thickarrayrulewidth
             \fi
      \ifnum0=`{\fi}}
\makeatother

\newlength{\thickarrayrulewidth}
\definecolor{DarkBlue}{HTML}{0f006e}

\usepackage{tikz}
\usetikzlibrary{arrows,fadings,patterns,shapes}
\usetikzlibrary{decorations.pathmorphing}
\usetikzlibrary{decorations.pathreplacing}
\usepackage{alphalph}
\usepackage{flafter}

\renewcommand{\div}{\operatorname{div}}
\newcommand{\grad}{\nabla}

\newcommand{\vv}[1]{\mathbf{#1}}

\newcommand{\filter}{h}
\newcommand{\inimage}{z}
\newcommand{\outimage}{\Hat{u}}
\newcommand{\targetimage}{u}

\newcommand{\comment}[1]{\textcolor{blue}{[\small{}#1]}}
\newcommand{\todo}[2][]{\comment{%
\textsc{Todo}\ifthenelse{\equal{#1}{}}{}{(#1)}: #2}}

\begin{document}

\supertitle{Stylization: Predefined and Personalized}
\title{Image Stylization: From Predefined to Personalized}

\author{
\au{Ignacio~Garcia-Dorado}, 
\au{Pascal~Getreuer},
\au{Bartlomiej~Wronski},
\au{Peyman~Milanfar}}

\address{\add{Google Research}}

\begin{abstract}

\looseness=-1 We present a framework for interactive design of new image stylizations using a wide range of predefined filter blocks. Both novel and off-the-shelf image filtering and rendering techniques are extended and combined to allow the user to unleash their creativity to intuitively invent, modify, and tune new styles from a given set of filters. In parallel to this manual design, we propose a novel procedural approach that automatically assembles sequences of filters, leading to unique and novel styles. An important aim of  our framework is to allow for interactive exploration and design, as well as to enable videos and camera streams to be stylized on the fly. In order to achieve this real-time performance, we use the \textit{Best Linear Adaptive Enhancement} (BLADE) framework --  
an interpretable shallow machine learning method that simulates complex filter blocks in real time. Our representative results include over a dozen styles designed using our interactive tool, a set of styles created procedurally, and new filters trained with our BLADE approach. 
\end{abstract}

\maketitle

\section{Introduction}

From the moment the camera was invented there has always been an interest to raise the bar of realism: capturing higher resolution images, inclusion, improvement, and precision of color, or even the addition of 3D scene depth. Fine art and photography have different goals when it comes to rendering a scene. While the former focuses on the aesthetic where the artist reflects their creative ideas using brush, paint, and a blank canvas, the latter aims to capture the intent of the photographer using a technical piece of equipment and (in more modern incarnations) software. \textit{Stylization} of photos is an approach that bridges the two worlds, allowing us to explore artistic expression on a canvas of already-captured images. Stylization creates evocative, abstract representations of natural and synthetic scenes, and is not limited to static photographs. It can also be applied to on video footage and video games as an additional dimension of artistic expression \cite{movie_vincent}.

Related to stylization is the concept of \textit{rotoscoping}, an animation technique used to trace over motion pictures footage, frame by frame, to produce realistic action. The technology dates back to the 1910s when Max Fleischer used it for cinematic storytelling. Besides the usage of classic cartoons like \textit{Popeye} and \textit{Betty Boop}, it garnered special attention with the '80s music video \textit{Take on Me}~\cite{song_take}. A few additional examples have been created since then, including \textit{A Scanner Darkly}~\cite{movie_dark} in the '90s, and the recent masterpiece \textit{Loving Vincent}~\cite{movie_vincent}. This technique allows the creation of stunning visual effects that enable the animator to dramatically changes the style. The main drawback is the manual labor.

In the early 90's, stylization was used by artists and designers to communicate, to abstract their ideas, and to express themselves. Technical illustration uses stylization to explain better the object's parts. By using the silhouettes and feature lines of the object it is easier to communicate concepts and remove spurious detail. The first example of stylization is arguably Haeberli~\cite{haeberli1990paint}. This seminal work achieved stylization by creating painterly images from a collection of brush strokes that were computed using local attributes of an image. This work was extended in the research community by many others (\cite{litwinowicz1997processing,kang2007coherent,kyprianidis2008image,kang2009flow,winnemoller2012xdog}). Automated stylization enabled practical application of stylization to video, for the first time used in the movie \textit{What Dreams May Come}~\cite{movie_dreams}. Since then, many movies (e.g., \textit{Waking Life}~\cite{movie_waking} and \textit{Sim City}~\cite{movie_sim_city}) and an increasing number of apps have included stylization of their content.

Nowadays, in the era of mobile photography, widely popular applications such as Instagram, Snapchat, Facebook, and Google Photos use stylization filters to alter the captured world. Millions of pictures are shared daily and in most cases the photos are processed with some filters. Since most pictures are stylized, the need to identify pictures that have {\em not} been altered has increased. Users often identify such pictures with the hash-tag `\textit{\#nofilter}' to show that no alterations were made (e.g., in Instagram there are over 260M photos posted with such a tag). We use stylization to express our emotions and feelings~\cite{hu2014we} and to increase the likelihood that our pictures are viewed and engaged with in social media~\cite{bakhshi2015we}. In this work we propose to extend the expressiveness of a photograph by allowing the user to create their own personalized stylizations.
Note that although this kind of stylization can be achieved with specialized tools such as Photoshop and Gimp, the required knowledge is out of reach of most users. Our goal is to bring user-customisable stylization to users of mobile devices, which requires reduced processing cost and simpler UI compared to desktop.
Our work will describe both a wide set of predefined stylizations, to the creation of unique stylizations tuned and designed by the user. 

\begin{figure}[tb]
\centering
\includegraphics[width=\linewidth]{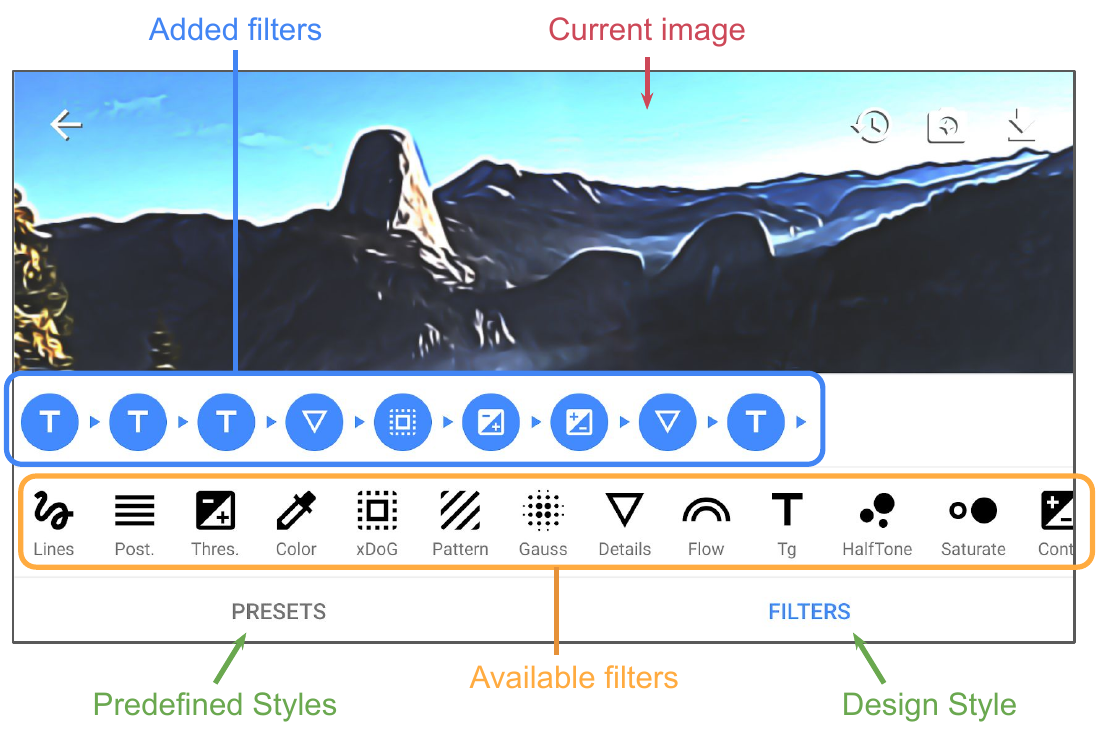}
\caption{\label{fig:fig_style_design_app}Experimental Design App. The users adds/modifies/replaces over 20 block filters to create a new stylization using a real time design application. A set of predefined styles are available as a starting point to be further customized by the user. See Sec.~\ref{subsec:interactive} for details.}
\end{figure}

We present an interactive design framework (see Fig.~\ref{fig:fig_style_design_app}) to create new stylizations using a wide range of predefined filter blocks. While the most common filters can be off-the-shelf image processing and rendering techniques, here we apply and combine them in novel ways. When designing the filter blocks (see Sec.~\ref{subsec:filters}), we recognize that the most interesting filters (e.g., Flow-XDoG) are too slow for interactive or real-time use. We expand the \textit{Best Linear Adaptive Enhancement} (BLADE) of Getreuer et al.~\cite{getreuer2018blade}, a lightweight, yet effective learning approach that is \emph{trainable} yet \emph{computationally simple} and \emph{interpretable}, to emulate a chosen set of filters in this stylization task. We show that using a shallow and easy to train machine learning method, we can add to our design toolbox a set of accurate approximations of significantly more complex and expensive filters. Our approach can be seen as a shallow two-layer network, where the first layer is predetermined and the second layer is trained. We show that this simple network structure allows for inference that is computationally efficient, easy to train, interpretable, and sufficiently flexible to perform the task of complex stylization.

\noindent The main contributions of this work can be summarized as follows:
\begin{itemize} 
\item We present an interactive design framework for stylization. The resulting tool allows for modifying, tuning, and designing new styles on the fly.
\item We extend BLADE, a light-weight and interpretable machine learning framework that is straightforward to train, to able to perform real-time inference of complex filters on mobile devices.
\item We propose a novel procedural style generation technique built on our design tool. Through simple rules we create original stylization effects from novel and automated combinations of filter blocks.
\end{itemize}

The rest of the paper is organized as follows: Section~\ref{sec:related_work} reviews previous work. Section~\ref{sec:overview} is an overview of our system. Section~\ref{sec:stylization} introduces our design application and the filter blocks. Section~\ref{sec:blade} explains how we use BLADE to achieve real time performance. Section~\ref{sec:stylization_design} presents our stylization design results. Finally, Section~\ref{sec:conclusions} contains conclusions and future work.

\section{Related work}\label{sec:related_work}

We cover related work in filtering and stylization research. Note that stylization of images and videos represents a broad research area, thus we review three main topics: filtering, video stylization, style transfer, and learnable filtering.


\subsection{Filtering and Abstraction}\label{subsec:stylization} 

Winnem{\"o}ller et al.~\cite{winnemoller2012xdog} proposed the use of eXtended Difference-of-Gaussians (XDoG) to create interesting sketch and hatching effects. Kang et al.~\cite{kang2007coherent} extended XDoG by adding an edge tangent flow block to create smooth edges.

Kyprianidis and D{\"o}llner~\cite{kyprianidis2008image} used oriented separable filters and XDoG to achieve a high level of image abstraction. Kang et al.~\cite{kang2009flow} further improved the level of abstraction by adding a flow-based step. Other more complex algorithms simplify images using advanced multi-scale detail image decomposition~\cite{talebi2016fast}.

For more general background on modern approaches to image filtering, we refer the reader to \cite{milanfar2013tour,milanfar2013symmetrizing}. For a more comprehensive survey of artistic stylizations we refer the reader to Kyprianidis et al.~\cite{kyprianidis2013state}. We employ some of these filters as building blocks of our stylization system as described in Section~\ref{sec:stylization}.

\subsection{Video Stylization}

As an extension of image filtering, some works have focused on speeding up the stylization process to run at interactive rates. Winnem{\"o}ller et al.~\cite{winnemoller2006real} used a bilateral filter iteratively to abstract the input, then quantize the background color and overlay XDoG to produce strong outlines. Our system can achieve similar results using a different set of filters. Barnes et al.~\cite{barnes2015patchtable} proposed to precompute a multidimensional hash table to accelerate the process of finding replacement patches. This structure enables them to stylize a video in real time using a large collection of patch examples

\subsection{Style Transfer} 

As an alternative to explicit filter creation, a wide range of works have developed a technique called \textit{style-transfer}. Style transfer is a process of migrating a style from a given image (reference) to the content of another (target), synthesizing a new image which is an aesthetic mixture of the two. Recent work on this problem uses Convolutional Neural Networks (CNN). Gatys et. al.~\cite{gatys2016image} posed the style-transfer problem as an energy minimization task, seeking an image close to the target using output of a CNN as a loss function. As an alternative to CNNs, Elad and Milanfar~\cite{elad2017style} presented an approach based on texture synthesis. Their approach copies patches from the reference image to the target while maintaining the main features of the content image using a hierarchical structure.

Despite the visual appeal of these approaches, their complexity is a major drawback. The method described by Gatys et. al.~\cite{gatys2017controlling} can take up to an hour to stylize a single image. More recent works have focused on addressing this issue. Johnson et al.~\cite{johnson2016perceptual} achieved real-time style transfer with simplified networks running on a high-end desktop GPU. However, achieving similar results on a full HD image on a mobile device would require tens of seconds. Elad and Milanfar's~\cite{elad2017style} approach takes minutes to run on a mobile device.

Another alternative is patch-based style transfer. Such methods transfer the style by finding and applying patches of the reference image in the target image. Barnes et al.~\cite{barnes2015patchtable} presented a method to efficiently query patches within a large dataset and replace each patch of the target image with one from the reference image. Friego et al.~\cite{frigo2016split} used local image features to determine the best scale of a patch. These approaches do not allow control over the color, line weight, and other features of stylization.

Style transfer has several other drawbacks. First, output quality depends directly on the reference image. While often considered an advantage, having a specific template can generate inconsistent and undesirable results for different inputs (e.g., very bright/dark images) or transform the content in areas in which we would like to preserve details. Second, current style transfer approaches do not provide sufficient aesthetic control. Gatys et. al.~\cite{gatys2017controlling} extended their own work to introduce control over color, scale, and spatial location. However, this does not allow the fine tuning required to design new stylizations. 

Note that our method does not compete against these approaches; each of the techniques above can be incorporated into our system as a new block to further enrich the creative toolbox. Moreover, note that our method focuses on the design aspect of new stylizations, and while we aim to create real-time filters that can create interesting results, we also want to allow for creating filters that can be tuned to a specific designer intent.

\subsection{Trainable Filters}\label{subsec:trainable_filters}

As previously discussed, we may express many methods of stylization as a cascade of filtering operations which are adaptive to image content in some way. Turning the problem on its head, we are able to discover and parameterise a sequence of filters which emulate the effects of those more complicated systems. These \textit{trainable filter} approaches include the recent Deep Learning approaches, variational methods, or closely connected methods in partial differential equations, Markov random fields, and maximum a posteriori estimation. A particularly successful direction is ``unrolling'' (or ``unfolding''), which is described generically for instance by Liu et al.~\cite{liu2018proximal}. The recipe is to formulate a task as an optimization, solve it with an iterative algorithm (e.g.\ with gradient descent or proximal methods), unroll several iterations, then substitute portions of the algorithm with trainable parameters. Compared to generic convolutional architectures, the advantage of this unrolling approach is that it tends to reduce the needed number of parameters, training data, and inference computational cost for a given level of quality. For instance
Chen and Pock's trainable nonlinear reaction diffusion~\cite{chen2017trainable} and Lefkimmiatis's work~\cite{lefkimmiatis2018universal} are image denoising networks designed by unrolling variational optimization algorithms. Besides denoising, deep unrolling has been applied to tasks such as image deblurring~\cite{liu2018proximal,corbineau2019learned,li2019algorithm}. While deep learning methods are capable of achieving impressive quality, they are hard to analyze and debug, and still too expensive to run interactively on smartphones at full-resolution.

For mobile-friendly filtering, we extend the Best Linear Adaptive Enhancement (BLADE) framework of Getreuer et al.~\cite{getreuer2018blade} (based on the RAISR method of Romano et al.~\cite{romano2017raisr}) to achieve real time inference of complex filters. For instance with $5\times 5$ filters on the Google Pixel 2018 phone, our CPU implementation runs at 38.21~MP/s and our GPU implementation at 223.03~MP/s.

\section{Overview}\label{sec:overview}
\begin{figure}[tb]
\centering
\includegraphics[width=\linewidth]{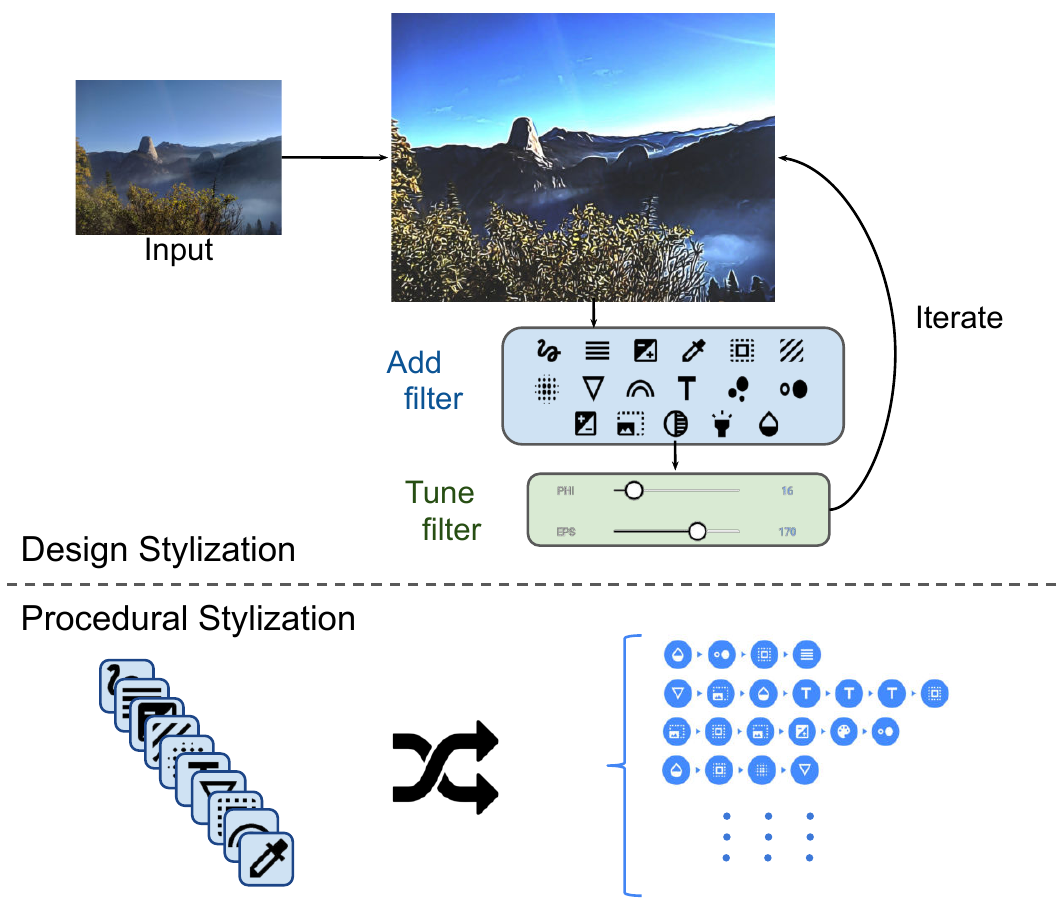}
\caption{\label{fig:2019_schematic}Diagram of our method: Stylization design and procedural generation.}
\end{figure}

In this section, we show an overview of our work.
Fig.~\ref{fig:2019_schematic} illustrates our two workflows for stylization design. 
In the first workflow (top) the user provides as input an image or uses a video live-stream (e.g., selfie camera stream) and manually selects filters and modifies parameters until the desired effect is achieved. In Sec.~\ref{sec:stylization} we present the interactive interface (Sec.~\ref{subsec:interactive}) and the filter blocks (Sec.~\ref{subsec:filters}).
In the second workflow (bottom) a style sequence is automatically generated by evaluating random combinations of filters and parameter sets. We call this \textit{procedural stylization}. In this case, style sequences are selected based on manual or automated subjective assessment (NIMA~\cite{talebi2018nima}, a no-reference aesthetic prediction network). 
 
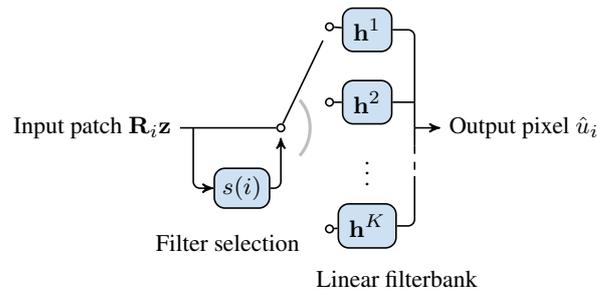
\begin{figure}[tb]
\centering
\mbox{%
\beginpgfgraphicnamed{images/inference_diagram}%
\input{images/inference_diagram.tikz}%
\endpgfgraphicnamed}
\caption{\label{fig:inference_diagram} BLADE inference.
$\vv{R}_i$ denotes extraction of a patch centered at pixel $i$. For a given
output pixel $\outimage_i$, we only need to evaluate the one linear filter that
is selected, $\filter^{s(i)}$.}
\end{figure}

In addition to relatively well-known linear filters (e.g., blurring and saturation) we include more sophisticated effects based on the use of BLADE filters (Fig.~\ref{fig:inference_diagram}) to enable real time inference on mobile devices. Sec.~\ref{sec:blade} explains the details of BLADE: inference~(\ref{subsec:inference}), training~(\ref{subsec:training}), filter selection with structure tensor features~(\ref{subsec:structure_tensor}), and real-time filtering~(\ref{subsec:real_time_filtering}).

\section{Stylization through filtering}\label{sec:stylization}

In this section, we describe an interactive tool we created for designing styles by combining filter blocks.

\subsection{Interactive Style Design}\label{subsec:interactive}

While most works (Sec.~\ref{subsec:stylization}) focus on creating one filter or a fixed set of filters to achieve a style, our goal is to create a flexible tool that allows anyone to design stylization filters, regardless of their technical ability.

The final result of our framework is as shown in Fig.~\ref{fig:fig_style_design_app}. The main components are:
\begin{enumerate}
\item A wide range of filter blocks (see Sec.~\ref{subsec:filters}).
\item The ability to tune the parameters for any filter through sliders.
\item Two layers: a black and white foreground layer used as alpha channel to display lines and contours; and a background layer for color. This separation provides added flexibility.
\item A visual flow diagram of filter blocks which allow any filter to be added, moved, reordered, removed, and tuned at any time. This is the essence of the system's interaction design.
\end{enumerate}

\subsection{Filter blocks}\label{subsec:filters} 

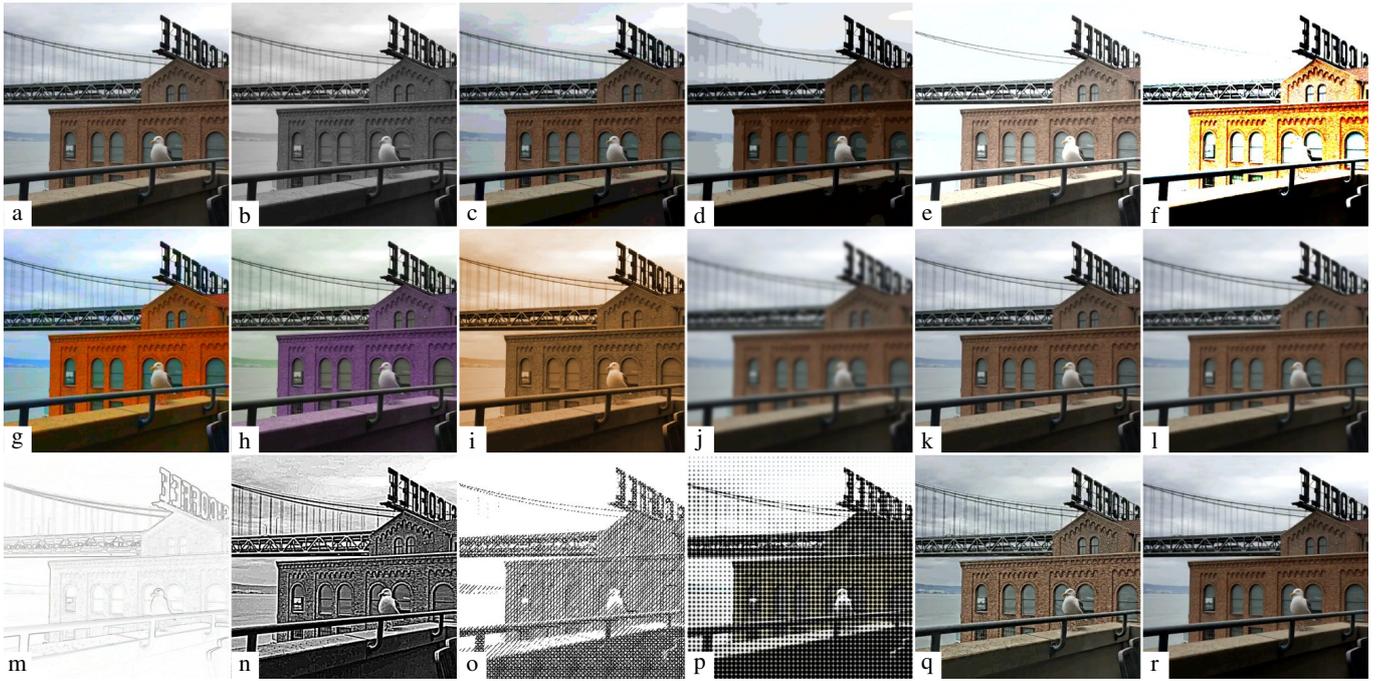
\begin {figure*}[hbt]
\centering
\begin{adjustbox}{width=\textwidth}
\begin{tikzpicture}
\node [inner sep=0pt,above right] (img) at (0,0){\includegraphics[width=\textwidth]{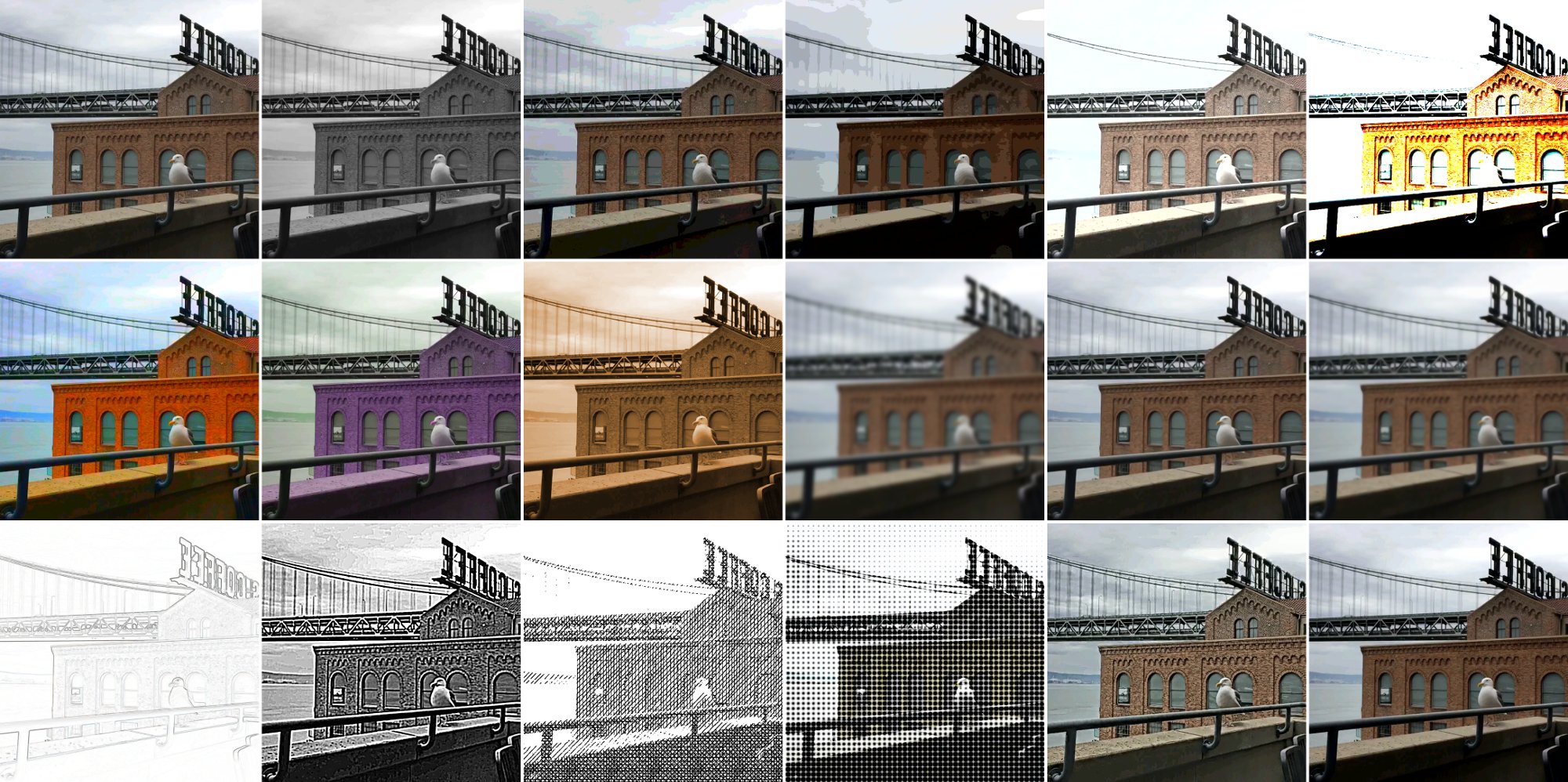}};
\begin{scope}[x=(img.south east),y=(img.north west)]
\foreach [count=\i] \y in {2,...,0}{
  \foreach [count=\j] \x in {0,...,5}{
     \draw [fill=white,white] (\x/6+0.0004,\y/3) rectangle (\x/6+0.02,\y/3+0.04);
     \draw (\x/6+0.01,\y/3+0.02) node[black] {\alphalph{\j+6*(\i-1)}};
          }}
\end{scope}
\end{tikzpicture}
\end{adjustbox}
\caption{\label{fig:all_filters}Filters: (a) Input, (b) To GrayScale, (c) Posterization, (d) Luma Posterization, (e) Brightness, (f) Soft Threshold, (g) Saturation, (h) Hue, (i) Colorize, (j) Gaussian Smoothing, (k) ETF, (l) TVF, (m) Sobel, (n) XDoG, (o) Pattern Filling, (p) Halftone, (q) Image Detail Control, and (r) Linear Equalization. See text for details.}
\end{figure*}

We implemented three categories of filter blocks: pixel operations, advanced filters, and histogram modification filters. The effect of each filter block on an example image is shown in Fig.~\ref{fig:all_filters}.

\vspace{4pt}
\noindent \textbf{A. Pixel-wise Operators:}
\vspace{-7pt}
\begin{itemize}
\item \textbf{To Grayscale:} Converts the image into grayscale/luminance and saves the chrominance channels ($\mathit{UV}$). This block is useful for applying any other block filter to just the luminance channel. [Fig.~\ref{fig:all_filters}~(b)].
\item \textbf{To Color:} Uses the current luma and converts the image back to $\mathit{RGB}$ using the $\mathit{UV}$ previously saved by the \textit{To GrayScale} block.
\item \textbf{Posterization:} Discretizes continuous image colors (e.g., 255 levels) to regions of fewer tones, e.g.\ $\mathit{levels} = 10$, [Fig.~\ref{fig:all_filters}~(c)].
\item \textbf{Luma Posterization:} Posterizes the image in the luma channel by converting the image to grayscale, applying \textit{Posterization}, and converting it back to color [Fig.~\ref{fig:all_filters}~(d)].
\item \textbf{Brightness:} Multiplies the luma channel by a user-selected $\mathit{brightness}$ constant, clipping the output values [Fig.~\ref{fig:all_filters}~(e)].
\item \textbf{Soft Threshold~\cite{winnemoller2012xdog}:} Computes the following function for each pixel
\begin{equation}\label{e:softthresh}
\mathit{output} = 1 + \tanh\bigl(\min(0, \phi \cdot (\mathit{input} - \epsilon))\bigr)
\end{equation}
where $\phi$ determines the slope and $\epsilon$ the cut-off. For grayscale images, this block behaves like a binary cut-off that preserves smooth transitions. For color images, it simplifies each $\mathit{RGB}$ channel into two levels [Fig.~\ref{fig:all_filters}~(f)].
\item \textbf{Saturation:} Makes the colors more vivid or more muted by adding or subtracting in $\mathit{RGB}$ the grayscale image tuned by a parameter [Fig.~\ref{fig:all_filters}~(g)].
\item \textbf{Hue:} Performs a color rotation in $\mathit{UV}$ space and adds a bias in $\mathit{RGB}$. This block is useful for changing the image's tint [Fig.~\ref{fig:all_filters}~(h)].
\item \textbf{Colorize:} Convert to monochrome using an $\mathit{HSL}$ palette transformation [Fig.~\ref{fig:all_filters}~(i)].\newline\newline 
\end{itemize}

\noindent \textbf{B. Spatial Filters:}
\vspace{-7pt}
\begin{itemize}
\item \textbf{Gaussian Smoothing:} Blurs and removes details and noise from the image with a controllable standard deviation ($\sigma$) parameter [Fig.~\ref{fig:all_filters}~(j)].
\item \textbf{Sobel Filter~\cite{sobel1990isotropic}:} Fast edge-detection filter [Fig.~\ref{fig:all_filters}~(m)].
\item \textbf{Scale:} Upscales or downscales the image using a user-selected scale parameter. This block can help to speed up computation (computing other blocks in a lower resolution) and alter the behavior of other scale-dependent filters.
\item \textbf{Pattern Filling Filter:} Uses \textit{Luma Posterization} to discretize the image into a set of levels, then, each pixel is replaced by a texture depending on its level. This block is useful for creating cross-hatching patterns [Fig.~\ref{fig:all_filters}~(o)].
\item \textbf{Halftone:} Replaces colors with a set of dots that vary in size and color. This style mimics the behavior of the four-color printing process traditionally used to print comics [Fig.~\ref{fig:all_filters}~(p)].
\end{itemize}

\noindent \textbf{C. BLADE Filters:}
\vspace{-7pt}
\begin{itemize}
\item \textbf{Edge Tangent Flow (ETF)~\cite{kang2007coherent}:} Creates an impressionistic oil painting effect. This method uses a kernel-based nonlinear smoothing of vector field inspired by bilateral filtering  [Fig.~\ref{fig:all_filters}~(k)].
\item \textbf{Total Variation Flow (TVF)~\cite{louchet2011total}:} Makes image piecewise constant with an anisotropic diffusion filter [Fig.~\ref{fig:all_filters}~(l)].
\item \textbf{Flow XDoG~\cite{winnemoller2012xdog}:} Uses \textit{Difference-of-Gaussians} to find the edges of the image. In our implementation, the user can control the variance of the main Gaussian ($\sigma$) and the multiplier ($p$) [Fig.~\ref{fig:all_filters}~(n)].
\item \textbf{Detail Control~\cite{talebi2016fast}:} Controls the details of the image by adding the residual of the image to its filtered version multiplied by a $\delta$. Setting $\delta < 0$ smooths the image while $\delta>0$ adds details [Fig.~\ref{fig:all_filters}~(q)]. 
\end{itemize}

\noindent \textbf{D. Histogram Operators:}
\vspace{-7pt}
\begin{itemize}
\item \textbf{Linear Histogram Equalization:} Equalizes the luma channel expanding the $p_{l}$ to zero and the $p_{h}$ to 255. Normally we choose $l=5$ and $h=95$ such that the $5\%$ percentile is moved to zero and the $95\%$ is moved to 255, thereby increasing the image's dynamic range [Fig.~\ref{fig:all_filters}~(r)].
\item \textbf{Histogram Minimum Dynamic Range:} Computes the percentile $5\%$ and $95\%$ on the luma histogram and expands (if necessary) the dynamic range to match the user parameter range $\mathit{DR}$.
\end{itemize}

These two filters are usually placed as the first filter in the pipeline to force the image to have a proper dynamic range to obtain a satisfactory result. For instance applying XDoG on a hazy or too bright/dark image would result in an almost entirely white output image.


\section{Best Linear Adaptive Enhancement}\label{sec:blade}

As we mentioned, the trade-off between quality and speed is a challenge with stylization. We want the freedom to apply elaborate techniques to produce interesting style effects, but on the other hand, computational efficiency is necessary to run efficiently. This is especially the case when processing full-resolution images or video as part of an interactive application on a mobile device. To this end, we leverage the BLADE framework to learn fast approximations to more complex filters.

Let $\vv{z}$ be an input grayscale image and $\vv{u}$ the target output grayscale image. We denote by subscript $z_i$ the $i$th pixel value at spatial position $i\in\Omega\subset\mathbb{Z}^2$. Let $\vv{h}^1,\ldots,\vv{h}^K$ denote a set of $K$ linear FIR filters, each having footprint or nonzero support $F\subset\mathbb{Z}^2$. The coefficients of these filters are learned.

\subsection{Inference}\label{subsec:inference}

We describe inference first to introduce the structure of the BLADE network architecture.
BLADE inference is a spatially-varying filter. For each output pixel, one filter in the bank is selected and applied:
\begin{equation}\label{e:blade_inference}
\outimage_i = \sum_{j\in F} \filter^{s(i)}_j \, \inimage_{i + j},
\end{equation}
where $s(i) \in \{1,\ldots,K\}$ denotes the index of the filter selected at the $i$th pixel (see filter selection at Sec.~\ref{subsec:structure_tensor}). Equivalently, inference (\ref{e:blade_inference}) can be written in vector notation as
\begin{equation}
\outimage_i = (\vv{\filter}^{s(i)})^T \vv{R}_i \vv{\inimage},
\end{equation}
where $(\cdot)^T$ denotes matrix transpose and $\vv{R}_i$ is the patch extraction operator defined by $(\vv{R}_i \vv{z})_j := z_{i+j}$, $j \in F$.

When computing $\Hat{u}_i$, only the selected filter needs to be evaluated. The complexity per pixel is $O(N)$ where $N = \lvert F \rvert$ is the number of pixels in the footprint. Notably, computation cost is independent of the number of filters $K$. Furthermore, inference is independent for each output pixel, so it is readily parallelized and implemented with high efficiency. 

We use features of the $2\times 2$ image structure tensor for the filter selection $s(i)$. This makes BLADE filtering adaptive to image edges and structure. In principle, BLADE can operate with any deterministic function of $\vv{z}$ as the selection rule.

To run BLADE on color images, we use the \textit{To~Grayscale} and \textit{To~Color} operations described earlier to extract the luma channel, filter luma with (\ref{e:blade_inference}), and reassemble a color image.

\subsection{Training}\label{subsec:training} 

To train the BLADE filters $\vv{h}^1, \ldots, \vv{h}^K$ to approximate an existing style effect, we first apply a (possibly slow) reference implementation of the effect to a set of images to create a training set of input image / target output image example pairs. We also train on $90^\circ$ rotations and flips of the images, augmenting the training set by a factor of 8.

For notational simplicity, we describe training for a single example image pair $\vv{z}$ and $\vv{u}$. The filters $\vv{h}^1, \ldots, \vv{h}^K$ are trained with a simple $L^2$ loss plus a quadratic regularization term,
\begin{equation}\label{e:training_objective}
\operatorname*{arg\,min}_{\vv{\filter}^1,\ldots,\vv{\filter}^K} \,
\sum_{k=1}^K \Bigl(\vv{(\filter}^k)^T \vv{Q} \vv{\filter}^k
+ \sum_{\substack{i\in\Omega:\\ s(i)=k}}
|\targetimage_i - (\vv{\filter}^k)^T \vv{R}_i \vv{z}|^2 \Bigr).
\end{equation}
The matrix $\vv{Q}$ determines the regularization. To encourage spatially-smooth filters, we set it to a discretization of $L^2$ norm of the filter's spatial gradient (~\cite{getreuer2018blade}). The inner sum is over the set of pixels $i$ where the $k$th filter is selected.

The training minimization (\ref{e:training_objective}) 
decouples over the filters, so each filter can be solved independently.

For each filter $\vv{h}$, its solution amounts to a multivariate linear regression with regularization. 
Denote by $\{i(1), \ldots,i(M)\}$ an enumeration of pixels where $s(i) = k$, and $M$ the number of such pixels. The optimal
filter $\vv{h}^k$ is
\begin{equation}
\vv{\filter} = (\vv{Q} + \vv{A}^T \vv{A})^{-1} \vv{A}^T \vv{b}
\end{equation}
where $A_{m,n} = (\vv{R}_{i(m)} \vv{z})_n$ and $b_m = \vv{u}_{i(m)}$. Rather than storing $\vv{A}$ and $\vv{b}$ themselves, it is possible to accumulate $\vv{A}^T \vv{A}$ as a matrix of size $N\times N$ and $\vv{A}^T \vv{b}$ as a length-$N$ vector. This way filters can be trained from any number of examples with a fixed amount of memory.

Once the filters $\vv{h}^1,\ldots,\vv{h}^K$ are trained, we visualize them as a tabular collage (Figs.~\ref{fig:etf_filters}, \ref{fig:fdog_filters}, \ref{fig:tv_flow_filters}, \ref{fig:details_-20_filters}). This is often an illuminating characterization of how BLADE will behave, as inference amounts to selecting among and applying these filters. For instance, a successful choice of filter selection mechanism $s(i)$ should allow the filters to specialize and take on diverse shapes, which we can inspect for visually. Additionally as described in detail in \cite{getreuer2018blade}, the variance of the regression residual is a good diagnostic to identify filters in need of more training examples or stronger regularization.

\begin{figure}[t]
\centering
\mbox{%
\beginpgfgraphicnamed{images/quantization}%
\input{images/quantization.tikz}%
\endpgfgraphicnamed}
\caption{\label{fig:quantization}Quantization of filter selection features with 
16 orientation bins, 5 strength bins, and 3 coherence bins.}
\end{figure}
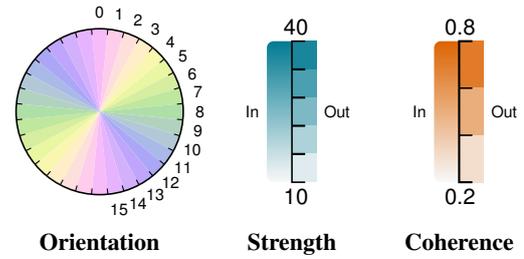

\subsection{Structure tensor features}\label{subsec:structure_tensor} 

Following \cite{romano2017raisr} and \cite{getreuer2018blade}, we make BLADE adaptive to the local image content by basing the filter selection $s(i)$ on features of the $2\times 2$ image structure tensor~\cite{forstner1987fast,bigun1987optimal,knutsson1993normalized}. In continuous space, the structure tensor is a $2\times 2$ matrix of spatial derivatives at every location:
\begin{equation}\label{e:structure_tensor}
J(\grad u) :=
\left(\begin{matrix} \partial_{x_1} u \\ \partial_{x_2} u
\end{matrix}\right)
\left(\begin{matrix} \partial_{x_1} u & \partial_{x_2} u
\end{matrix}\right),
\end{equation}
where above, $u(x)$ is a differentiable continuous-domain image,
$\partial_{x_1}$ and $\partial_{x_2}$ denote partial derivatives, and 
$\grad = (\partial_{x_1}, \partial_{x_2})^T$ denotes gradient.
In implementation the derivatives can be discretized with finite differences. We consider it important that the two spatial components of the gradient estimate are aligned with one another in order to avoid asymmetrical behavior in the processing. Using the smallest stencil with this property, we compute finite differences in $45^\circ$ rotated coordinates $x_1'$, $x_2'$ over a $2\times 2$ stencil as
\begin{align}
\begin{aligned}
\lefteqn{\tfrac{1}{\sqrt{2} h} \bigl(u(x_1 + 1, x_2) - u(x_1,x_2 + 1)\bigr)} \\
&= \partial_{x_1'} u(x_1+\tfrac{1}{2},x_2+\tfrac{1}{2}) + O(h^2),
\end{aligned} \\
\begin{aligned}
\lefteqn{\tfrac{1}{\sqrt{2} h} \bigl(u(x_1 + 1, x_2 + 1) - u(x_1,x_2)\bigr)} \\
&= \partial_{x_2'} u(x_1+\tfrac{1}{2},x_2+\tfrac{1}{2}) + O(h^2).
\end{aligned}
\end{align}
Next, each component of the structure tensor is spatially filtered with a Gaussian kernel $G_\rho$ with standard deviation $\rho$, 
\begin{equation}
J_\rho(\grad u) := G_\rho * J(\grad u).
\end{equation}
The filtered structure tensor $J_\rho(\grad u)$ is at each location an aggregate of the image statistics of its neighborhood, with the neighborhood size determined by $\rho$. These statistics robustly capture the predominant edge orientation and other local image characteristics, as studied for instance by Weickert~\cite{weickert1998anisotropic}, Zhu and Milanfar~\cite{zhu2009no}, and Takeda et al. \cite{takeda2006ICIP}.

At the $i$th pixel, we compute three features from the eigensystem of the $2\times 2$ filtered matrix $J_\rho(\grad u)_i$: (1) \textbf{orientation} as the angle of the dominant eigenvector, (2) \textbf{strength} as the square root of the dominant eigenvalue, and (3) \textbf{coherence} according to 
\begin{equation}
\text{coherence} = \frac{\sqrt{\lambda_1} - \sqrt{\lambda_2}}{
\sqrt{\lambda_1} + \sqrt{\lambda_2}}
\end{equation}
where $\lambda_1 \ge \lambda_2 \ge 0$ are the eigenvalues. 

To perform filter selection $s(i)$, we quantize these features to uniform bins, and we then view the bin indices as a three-dimensional index into the bank of filters. Fig.~\ref{fig:quantization} illustrates a typical quantization.

\subsection{Real-Time Advanced Filtering} \label{subsec:real_time_filtering} 

The following sections describe how we take advanced, computationally demanding 
effects and apply visually accurate approximations of them in real time using the BLADE
framework.

\begin{figure}[tb]
\centering
\includegraphics[width=\linewidth]{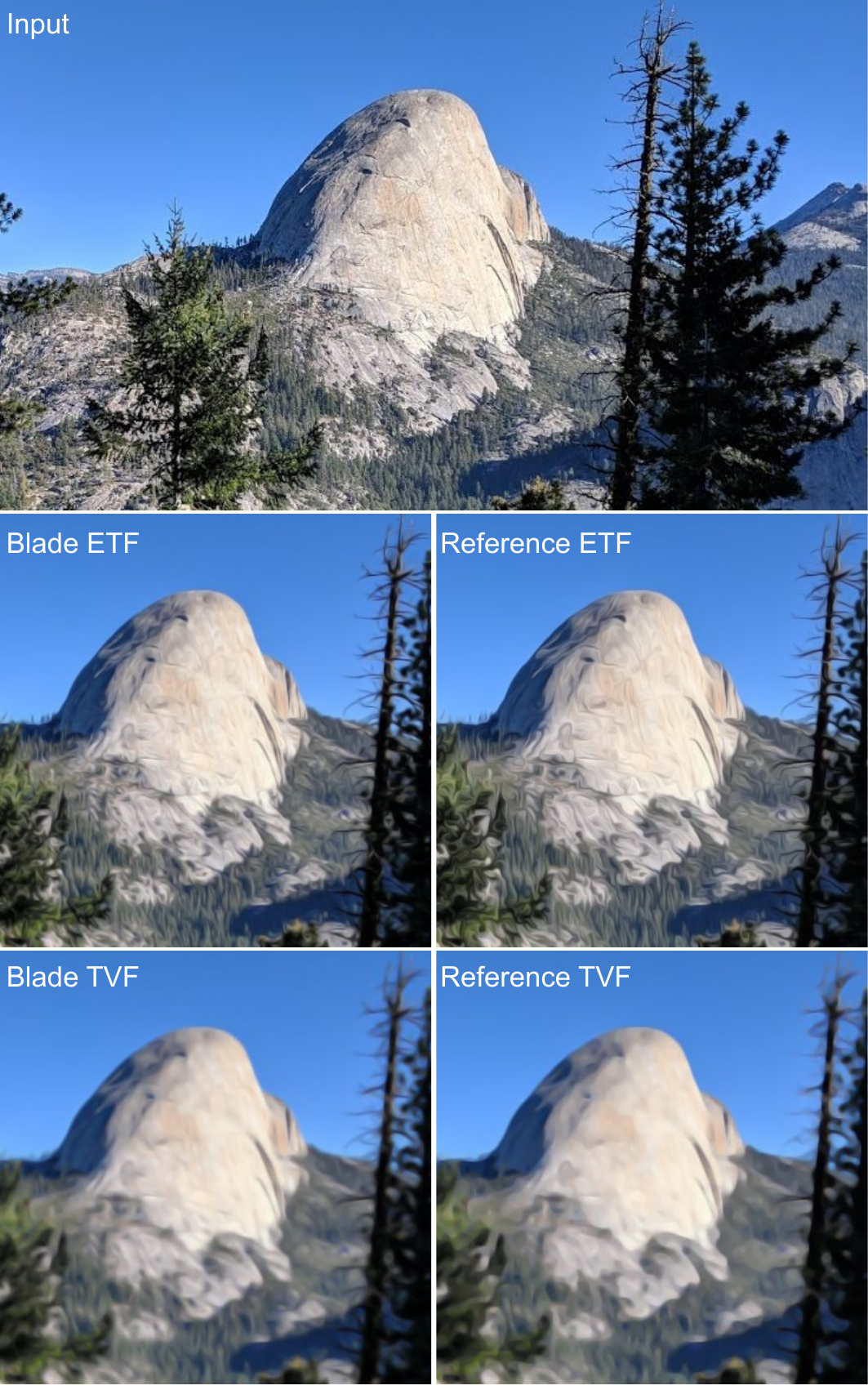}
\caption{\label{fig:blade_tg_tv}Example of approximated edge tangent flow and TV flow. BLADE has PSNR 34.85 dB and MSSIM 0.9619 for ETF and PSNR 35.86 dB and 0.9683 for TVF.}
\end{figure}

\begin{figure}[tb]
\centering
\includegraphics[width=\linewidth]{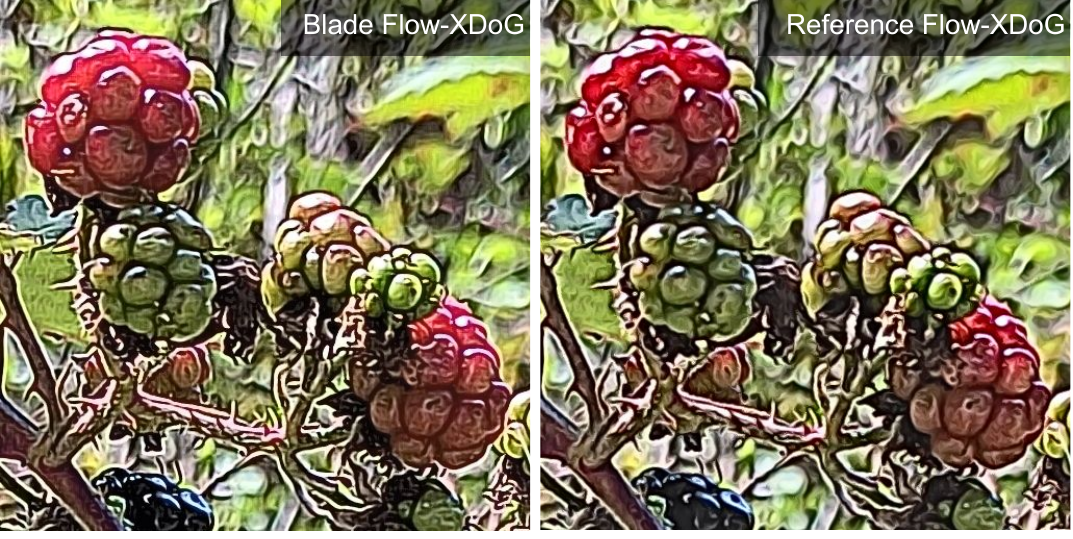}
\caption{\label{fig:blade_xdog}Example of approximated Flow-XDoG. Readers are encouraged to zoom in aggressively (200\% or more).}
\end{figure}
\begin{figure}[tb]
\centering
\includegraphics[width=\linewidth]{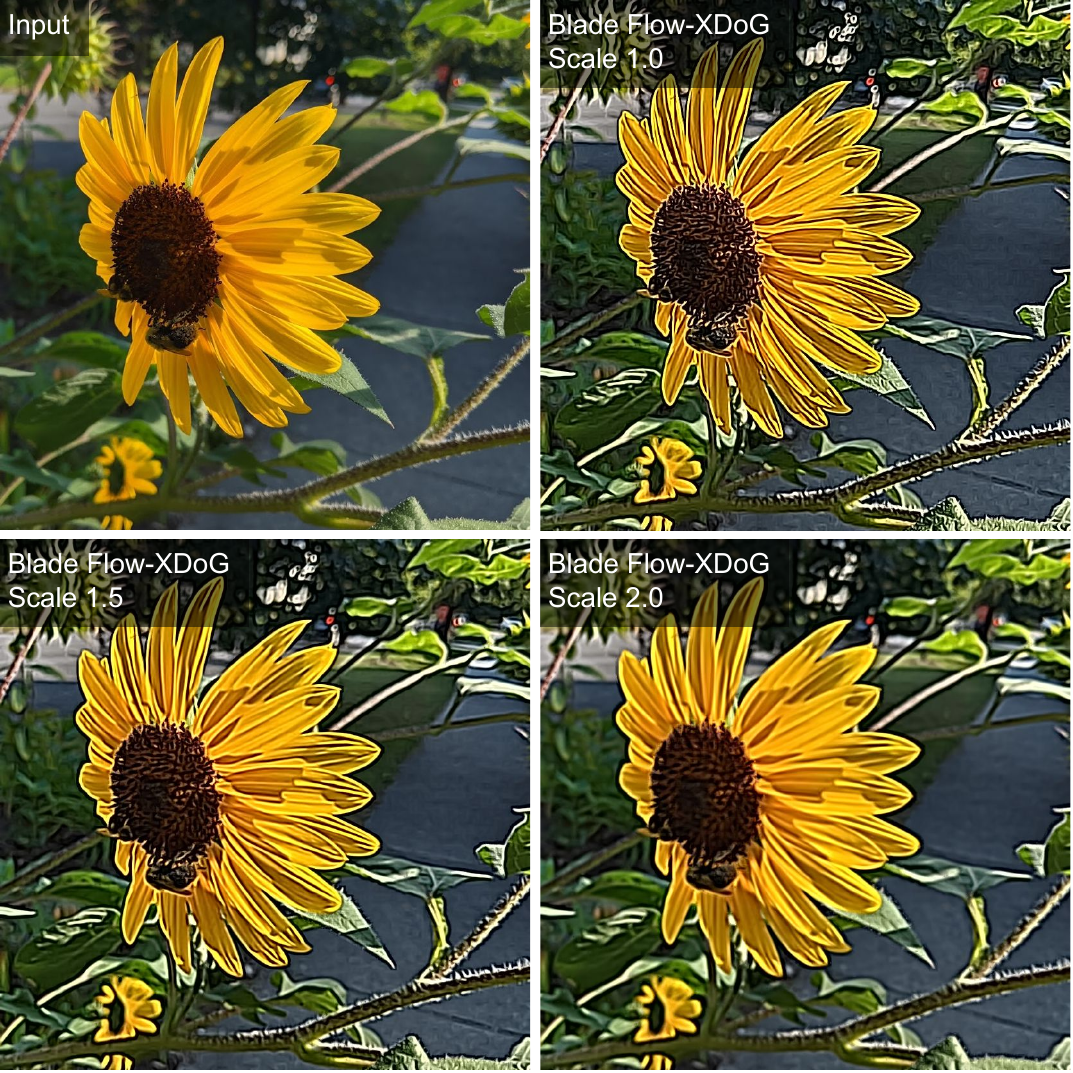}
\caption{\label{fig:blade_xdog_2x2}Examples of three scales of trained Flow-XDoG. Readers are encouraged to zoom in aggressively (300\% or more).}
\end{figure}

\begin{figure}[tb]
\centering
\includegraphics[width=\linewidth]{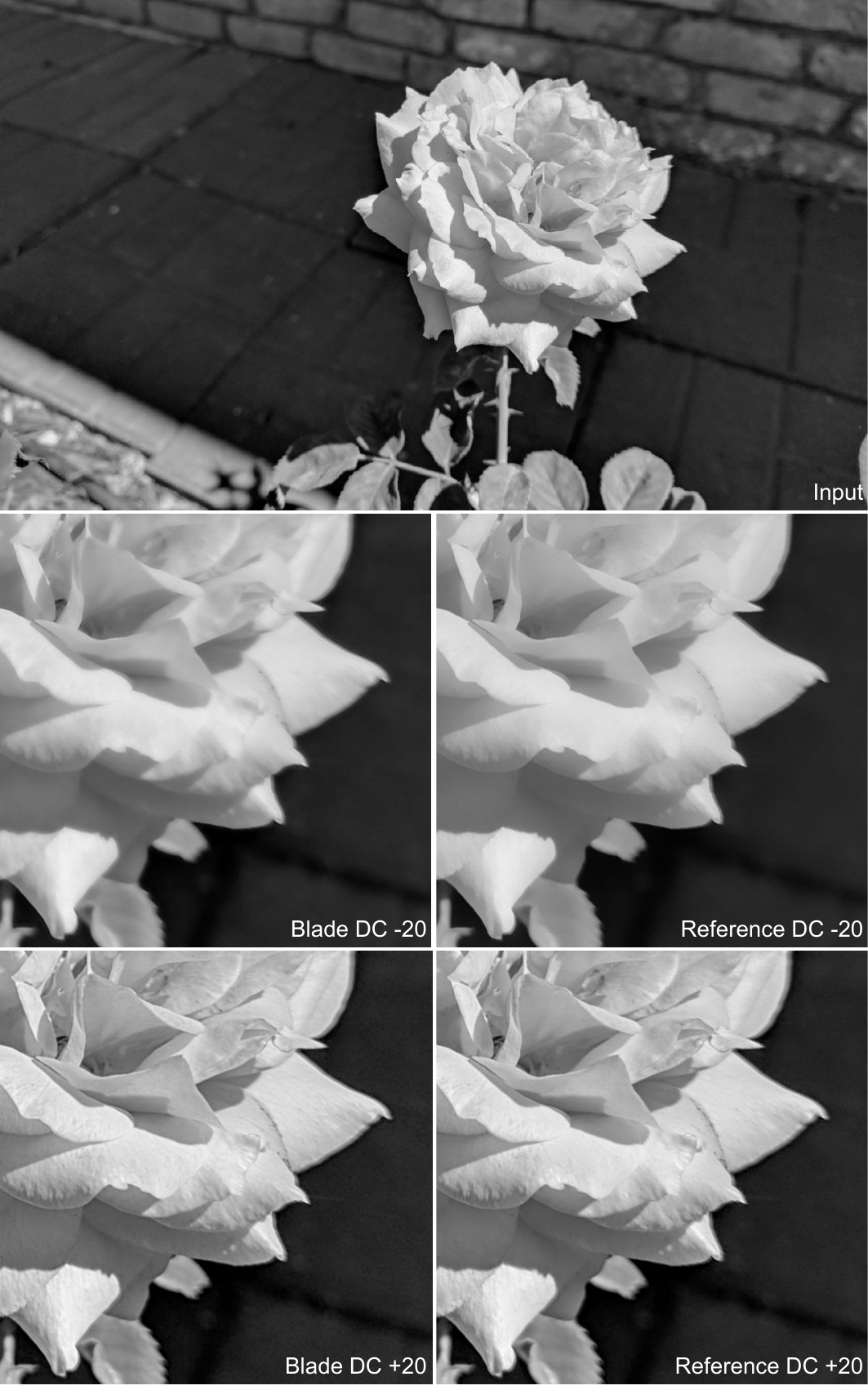}
\caption{\label{fig:blade_detail_control}Example of approximated \textit{Detail Control} for values \textit{-20} and \textit{+20}. BLADE has PSNR 38.38 dB and MSSIM 0.9845 for \textit{-20} and PSNR 41.36 dB and 0.9895 for \textit{+20}.}
\end{figure}

\subsubsection{TV flow}
Total variation (TV) flow is a classic anisotropic diffusion process that tends to regularize the image into piecewise-constant regions. TV flow flattens textures and details while retaining strong edges, which is aesthetically interesting in designing style effects for image simplification and producing a cartoon-like look.

In order to create a BLADE approximation of TV flow, we use as training target the modified TV flow definition of Marquina and Osher~\cite{marquina2000explicit},
\begin{equation}
\partial_t u = |\grad u| \div(\grad u /|\grad u|),
\end{equation}
in which $u(x)$ is a continuous-domain image.
Compared to the usual TV flow, this definition has an extra $|\grad u|$ factor, which mitigates 
the tendency to produce artificial edges in smooth gradients.

For training, we use a small private dataset of 32 photographs of size $1600\times 1200$ as input images. As a note, we find that the choice of input images does not have an impact, since BLADE does not have enough parameters (e.g., a 5x5 BLADE filter has 5400 parameters) to overfit the training data. To create the training target images, we apply the second-order scheme developed in \cite{marquina2000explicit} as the reference implementation to generate TV flow target images. For filter selection, we use 16 orientations, 4 strength bins, and 4 coherence bins ($K = 256$ filters total) and use filters of size $7\times 7$ (Fig.~\ref{fig:tv_flow_filters}).

Fig.~\ref{fig:blade_tg_tv} shows an example, comparing the input, TV flow reference implementation, and the fast BLADE approximation.

\subsubsection{Edge tangent flow}

Edge tangent flow (ETF) is another anisotropic diffusion equation with useful qualities for style effects. This process smooths the image along edges but not across them, as its name suggests. When applied to a photographic image, ETF tends to alter textures into flows and swirls like an impressionist-style oil painting. When applied as post processing to an edge mask (from the Sobel or Flow XDoG filter blocks), ETF makes lines visually more organic and flowy like hand-drawn strokes.

ETF anisotropic diffusion is defined mathematically by
\begin{equation}\label{e:etf_pde}
\partial_t u = \div\bigl(D(u) \grad u\bigr),
\end{equation}
where at each spatial location, $D(u)(x)$ is the $2\times 2$ outer product of the unit-magnitude local edge tangent orientation. The edge tangent orientation is obtained as the weaker eigenvector of the smoothed image structure tensor.

For training a BLADE approximation of ETF, we use the same dataset of 32 photographs, and use line integral convolution evolved with second-order Runge--Kutta as a reference implementation to generate ETF target images. We use 24 orientations and 3 coherence bins. We omit the strength feature, since ETF is essentially a one-dimensional diffusion at every point in the edge tangent orientation and this behaves irrespective of the gradient magnitude. We train filters of size $5\times 5$ (Fig.~\ref{fig:etf_filters}). Our CPU implementation on a Xeon E5-1650v3 runs BLADE ETF at 27.1 MP/s. Fig.~\ref{fig:blade_tg_tv} shows an example.

BLADE ETF has two interesting parameters: increasing the structure tensor smoothing parameter $\rho$ produces broader brush strokes, and applying multiple passes of the filter (more time steps of the diffusion equation) results in longer strokes.

\subsubsection{Flow-XDoG}

Extracting image edges is a fundamental element for line drawing or cartoon-like styles. We briefly review the origin of the Flow-XDoG filter:
\begin{itemize}
\item  The classical Marr--Hildreth approach to edge detection is Laplacian filtering, which can be made more noise robust by using a Laplacian of Gaussian filter. The DoG filter $G_{k\cdot\sigma} - G_\sigma$ with $k = 1.6$ is a close approximation to the Laplacian of Gaussian~\cite{marr1980theory}.
\item Improving on DoG, Kang et al.~\cite{kang2007coherent} introduced flow-based difference-of-Gaussians (FDoG). A basic implementation of FDoG is simply ETF followed by DoG filtering, which we could realize as two filter blocks. However, as Kang et al.\ develop, the edge tangent flow field can be used in a ``flow-guided'' DoG filter to obtain a cleaner output that responds more strongly on true edges.
\item In a later work, Winnem{\"o}ller et al.~\cite{winnemoller2012xdog} introduced extended difference-of Gaussians (XDoG). XDoG substitutes the 
highpass DoG filter $G_{k\sigma} - G_{\sigma'}$ with a high-emphasis filter $(1 + p) G_{k\sigma} - p G_\sigma$, with emphasis according to parameter $p$. Second, XDoG follows the filter the \textit{Soft Threshold} function (\ref{e:softthresh}). XDoG can be combined with FDoG, which we refer to as \emph{Flow-XDoG}.
\end{itemize}

To make a fast approximation of Flow-XDoG, we decompose it to two stages. First, BLADE is used to approximate a flow-guided version of the high-emphasis DoG filter $(1 + p) G_{k\sigma} - p G_\sigma$ (for which we use second-order Runge--Kutta line integral convolution as the reference implementation), and second, the soft threshold is applied. Being a simple pointwise operation, the soft threshold is straightforward to implement separately from BLADE, and this has the advantage that the soft threshold parameters $\phi$ and $\epsilon$ are tunable at inference time.

We use 16 orientations, 5 strength bins, and 3 coherence bins, and we train $7\times 7$ filters (Fig.~\ref{fig:fdog_filters}). Fig.~\ref{fig:blade_xdog} shows an example, comparing the input, Flow-XDoG reference implementation, and the fast BLADE approximation. Fig.~\ref{fig:blade_xdog_2x2} shows how we can change the scale of the image at training to produce a more subtle behavior of Flow-XDoG.

\subsubsection{Detail Control}

Smoothing or enhancing details is a versatile tool for stylization. Smoothing can help to remove the details to create an abstraction or it can enhance the details to emphasize the different aspect of the images. 
An effective detail control filter is Talebi and Milanfar's multilayer Laplacian enhancement~\cite{talebi2016fast}.

For training a BLADE approximation, we use the same dataset of 32 photographs, and use
the algorithm described in \cite{talebi2016fast} as a reference implementation to create training targets. 

We use 16 orientations, 5 strength bins, and 3 coherence bins, and we train $9\times 9$ filters (Fig.~\ref{fig:details_-20_filters}). Note this change affects only the luminance channel, therefore Fig.~\ref{fig:blade_detail_control} shows an example of control to smooth or increase the detail of that channel.

\section{Style Design}\label{sec:stylization_design}

In this section, we present the following results: a set of new styles generated by designers and a set of styles generated procedurally.

\begin{table}[tb]
\caption{\label{tab:computation_time}Style computation time of Fig.~\ref{fig:2019_main_1} and \ref{fig:2019_extra_styles} (in megapixels per second) \newline for \textit{`Device'} (Pixel 2018) and Desktop GPU.}
\begin{minipage}{.5\linewidth}
\small
\centering
\begin{tabular}{ccc}
\thickhline
\multicolumn{1}{l}{\multirow{2}{*}{Style}} & \multicolumn{1}{l}{\multirow{2}{*}{\begin{tabular}[c]{@{}c@{}}Device\\ MP/s\end{tabular}}} & \multicolumn{1}{l}{\multirow{2}{*}{\begin{tabular}[c]{@{}c@{}}Desktop\\ MP/s\end{tabular}}} \\ \\  \hline

1     & 124.4 & 1719.4 \\
2     & 156.9 & 2405.5 \\
3     & 128.4 & 1469.1 \\
4     & \hphantom{0}30.3 & \hphantom{0}357.5  \\
5     & \hphantom{0}92.3 & 1469.4  \\
6     & \hphantom{0}40.0 & \hphantom{0}458.7  \\ \thickhline
\end{tabular}
\end{minipage}
\begin{minipage}{.5\linewidth}
\centering
\begin{tabular}{ccc}
\thickhline
\multicolumn{1}{l}{\multirow{2}{*}{Style}} & \multicolumn{1}{l}{\multirow{2}{*}{\begin{tabular}[c]{@{}c@{}}Device\\ MP/s\end{tabular}}} & \multicolumn{1}{l}{\multirow{2}{*}{\begin{tabular}[c]{@{}c@{}}Desktop\\ MP/s\end{tabular}}} \\ \\  \hline
7     & 131.0  & 1637.5 \\
8     & 169.4  & 2295.9 \\
9     & 114.0  & \hphantom{0}675.2  \\
10     & 241.8  & 4069.9 \\
11     & 240.7 & 4138.8\\
12     & 126.0 & 1804.9 \\
13     & 101.3 & 1323.3 \\ \thickhline
\end{tabular}
\end{minipage}
\end{table}

\begin{figure*}[p]
\centering
\includegraphics[width=\linewidth]{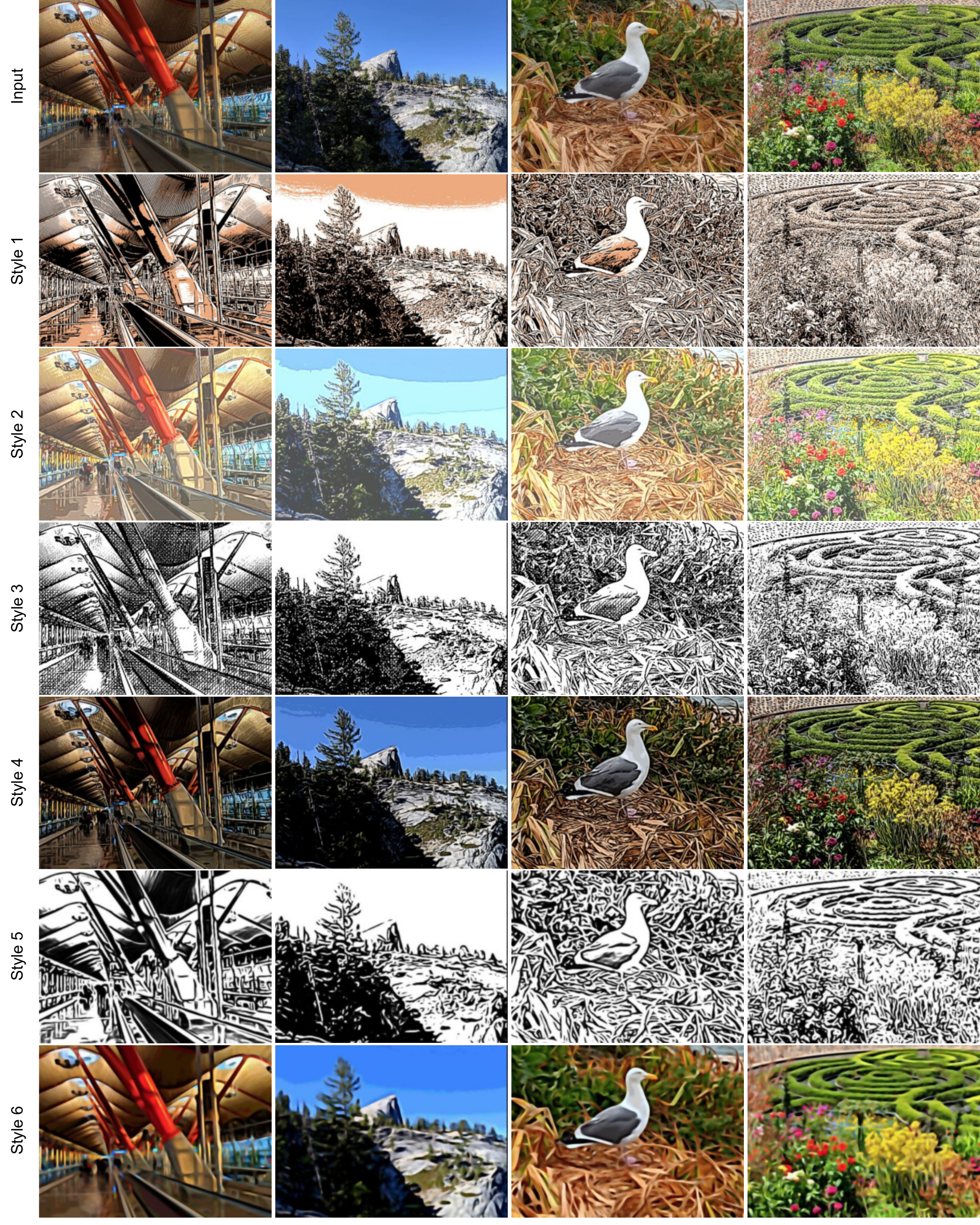}
\caption{\label{fig:2019_main_1}Input image on top and six different stylizations created with our tool. See details in Sec.~\ref{subsec:new_stylizations}. Readers are encouraged to zoom in aggressively (200\% or more).}
\end{figure*}

\begin{figure*}[tb]
\centering
\includegraphics[width=\linewidth]{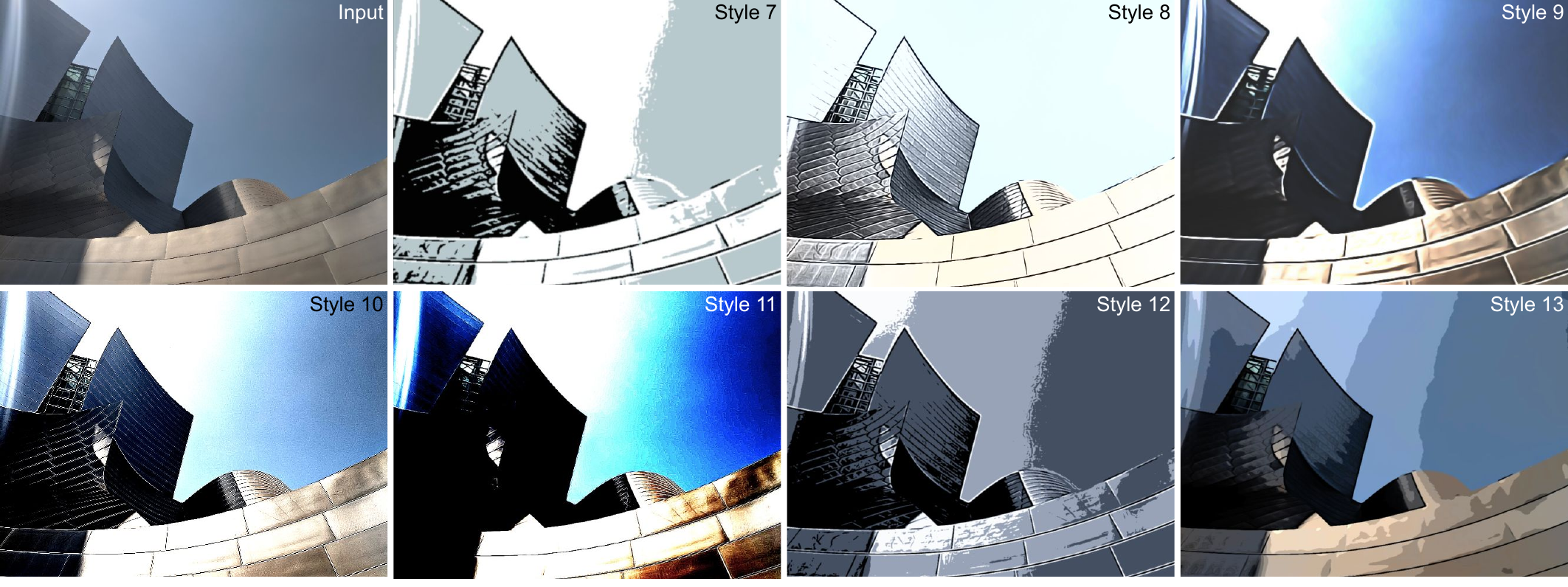}
\caption{\label{fig:2019_extra_styles}Additional stylization created with our tool. Readers are encouraged to zoom in aggressively (200\% or more).}
\end{figure*}


\subsection{Interactive Style Design}\label{subsec:new_stylizations}
We conducted several style design sessions with graphic designers who generated manually dozens of styles. Fig.~\ref{fig:2019_main_1} shows six styles and Fig.~\ref{fig:2019_extra_styles} shows an additional seven styles to give the reader some appreciation of the breadth of effects that could be deployed with this system.

Fig.~\ref{fig:2019_progress} depicts the design of each stylization from the input (left) to the final result (right) for six styles and Fig.~\ref{fig:2019_filters_2} shows a summary of how the styles are generated. 
\begin{itemize}
    \item \textbf{Style 1:} This style uses Flow-XDoG to transform the input then threshold and color are applied. Gaussian block is used to smooth the filtering. Here we present it in an orange tint but we created explored versions with dark blue, green, blue, and grayscale.
    \item \textbf{Style 2:} This style tries to imitate crayons. This stylization uses a smoothed and thresholded version of the image to apply a downscaled version of Flow-XDoG in grayscale, finally, the colors are saturated.
    \item \textbf{Style 3:} This style tries to create the effect of a sketch. Instead of using XDoG (as it was done in~\cite{winnemoller2012xdog}), we simplify this filter by using thresholding and apply a five-level hatching texture to the grayscale. Finally, we overlay lines to highlight contours.
    \item \textbf{Style 4:} This style tries to abstract or simplify the input. To achieve such a result, we remove details, downscale, and smooth with the watery ETF filter, we then apply Flow-XDoG and apply posterization.
    \item \textbf{Style 5:} This is a heavily stylized result. Details are removed and Flow-XDoG is applied and several ETV and TVF are applied. To get  the final output, we posterize the result and apply lines to highlight the edges.
    \item \textbf{Style 6:} We call this stylization \textit{blob}, it is a colorful abstraction of the image. To create it, we apply Flow-XDoG in its lowest scale (i.e., it smoothes the input), we posterize the result and remove details.
\end{itemize}

We show the performance on device and on desktop (Table~\ref{tab:computation_time}). The table shows that our method, even when using heavy filters (such as three times ETF), can achieve real-time rendering performance. When a style uses the basic set of filters it is possible to process a Full HD input video in under 1ms, when more advanced filters are used we achieve real time rendering over 30fps at viewfinder resolution. Desktop performance is one order of magnitude faster and we can achieve 16K video processing with over 60fps. 

\subsection{Procedural Styles}

\begin{figure}[tb]
\centering
\includegraphics[width=\linewidth]{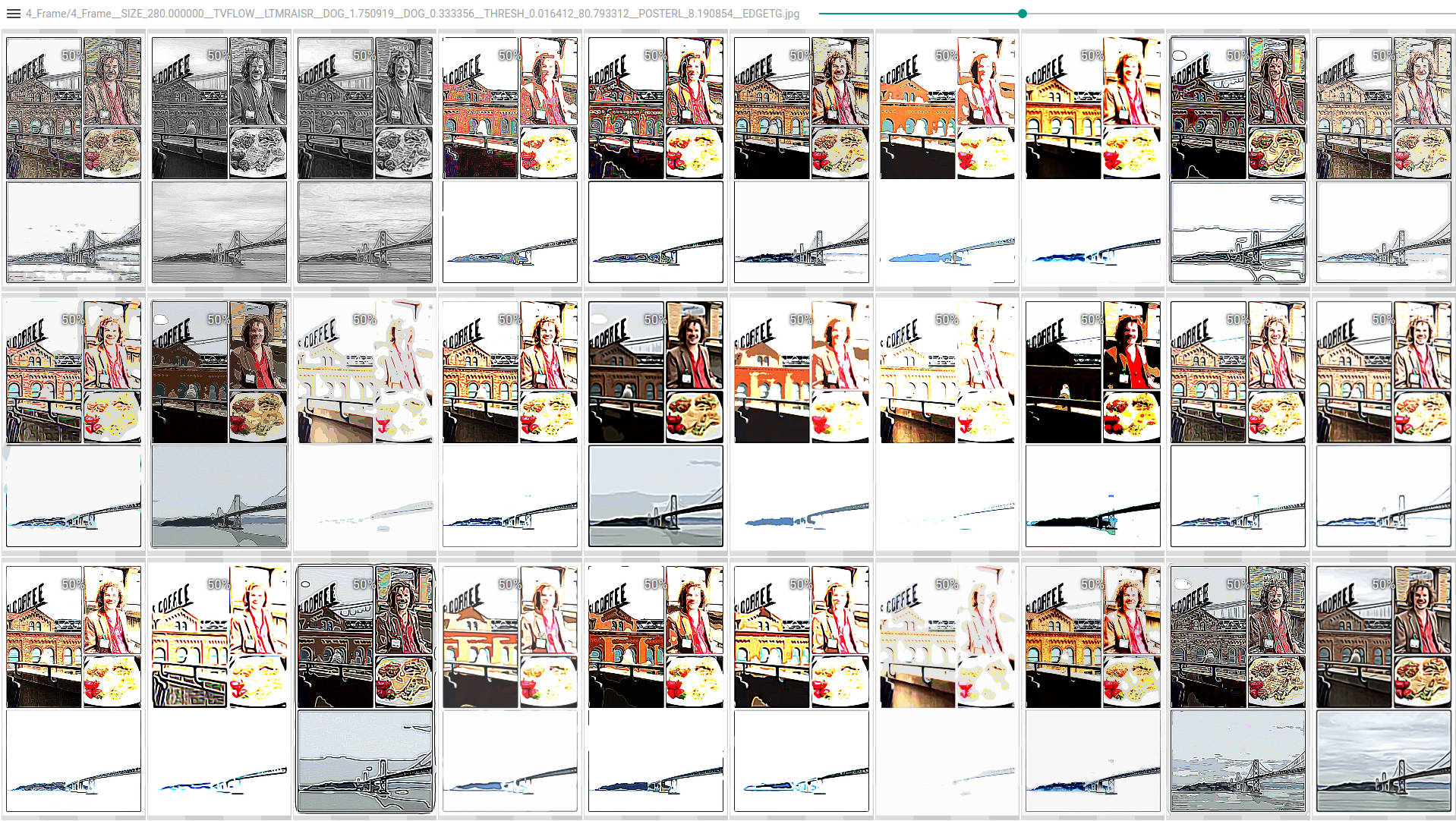}
\caption{\label{fig:exploration_tool}Visualization tool to explore alternative procedurally generated styles. Many alternatives (as seen in the figure) are not very interesting or distinct, but the system's real-time speed and created visualization tool makes it fast and easy to explore and identify promising alternatives.}
\end{figure}

\begin{figure*}[tb]
\centering
\includegraphics[width=\linewidth]{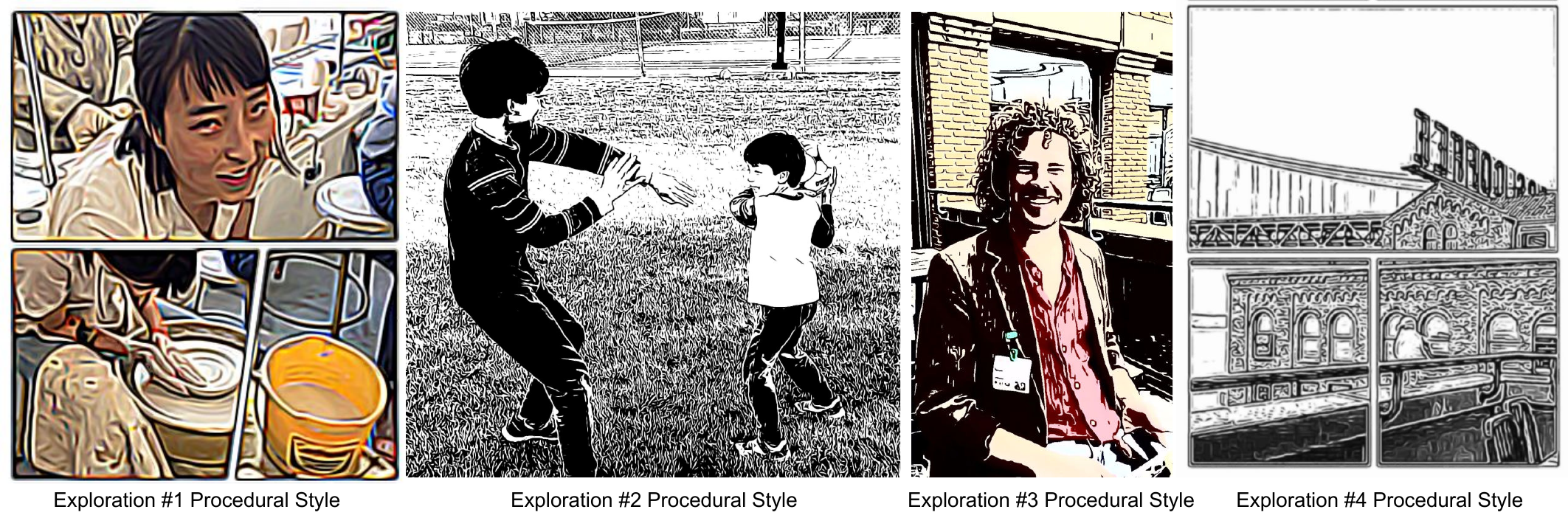}
\caption{\label{fig:pm_exploration}Manual exploration of procedurally generated styles. Readers are encouraged to zoom in aggressively (200\% or more).}
\end{figure*}

\begin{figure*}[tb]
\centering
\includegraphics[width=\linewidth]{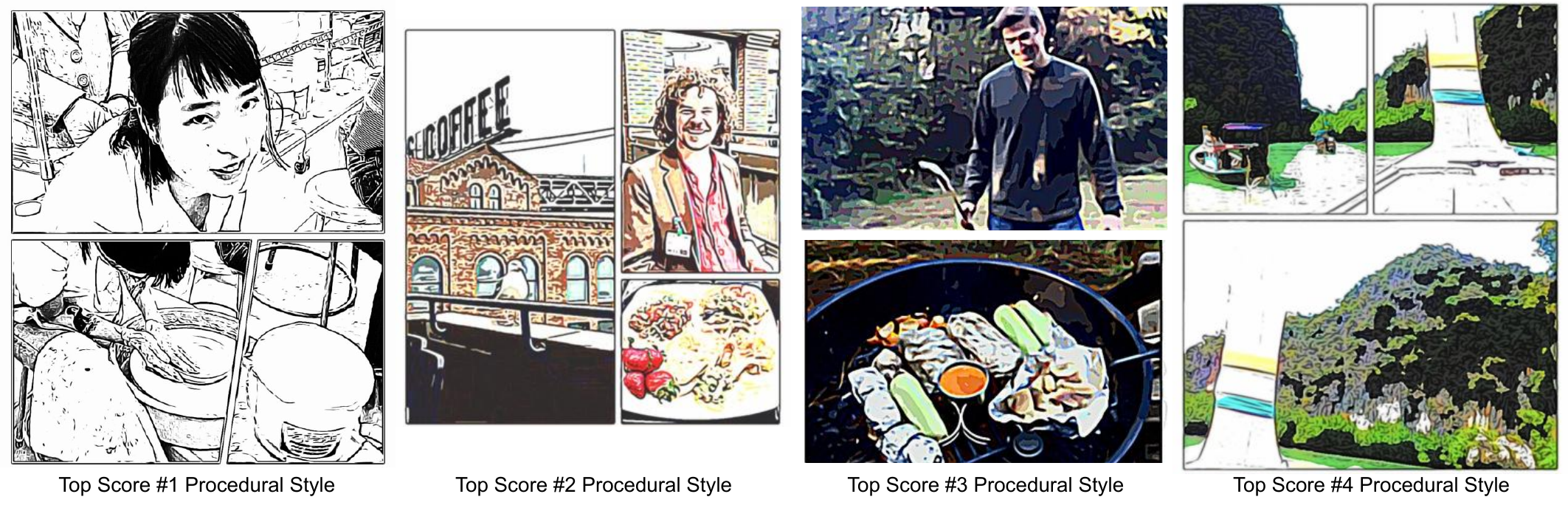}
\caption{\label{fig:pm_scored}Top-scored procedurally generated styles. Readers are encouraged to zoom in aggressively (200\% or more).}
\end{figure*}

Procedural modeling is technique that enables the generation of hundreds or thousands of examples using a limited set of rules. In the computer graphics community, the technique has been used to generate buildings, cities, trees, and more complex models; see for instance \cite{muller2006procedural,nishida2016interactive}. Based on this idea, we call \textit{procedural styles} those created randomly using a set of simple rules.

The simple set of rules for this experiment was to add a random number of filters between 4 and 9, with random input parameters: XDoG ($\sigma \in [0.5, 8.0]$ and $p \in [1, 40]$), TVF, Soft Threshold ($\phi \in [0.013, 0.059]$ and $\epsilon \in [50, 110]$), Detail Control ($\delta \in [-100, 60]$), Luma Posterization ($\mathit{level} \in [5, 12]$), Saturation ($\mathit{saturation} \in [1.5, 2.2]$), Size ($\mathit{size} \in [100, 300]$, and To GrayScale ($20\%$ probability). We also enforce a rule that XDoG and TVF are the only filters that can be added more than once (duplicating the rest of filters has the same effect as selecting different parameters).

We developed a web-based visualization tool to quickly review the procedurally generated alternatives (Fig.~\ref{fig:exploration_tool}). Fig.~\ref{fig:pm_exploration} shows four examples of styles found manually in under five minutes using the visualization tool. In order to automate this process, we used the Neural Image Assessment (NIMA) network of Talebi and Milanfar~\cite{talebi2018nima}, which evaluates the aesthetic quality of an image. Using this approach, we discovered the styles in Fig.~\ref{fig:pm_scored}.

\subsection{Computational Performance}\label{subsec:computational_performance}

One of the key goals of our system is allowing for interactive exploration and design of different styles and options, as well as being able to process videos and camera streams. 
Therefore, we have aimed to achieve real-time computational performance and decided for a GPU implementation.
Most image filters are standard image processing operations, and we have implemented them using OpenGL and GLSL shading language.

In the case of BLADE filtering, we have taken advantage of the fact that it is a fully parallelizable algorithm, and also implemented it using GPU GLSL shading language.
We have decoupled two main steps of the algorithm as separate full-image passes: filter bucket hash computation and applying of the selected filter.
Both passes run only on a single image channel (luminance) and take advantage of accelerated instructions and OpenGL extensions like \textit{ARB\_texture\_gather} that allow to process four pixels at the same time.
To optimize the bucket index computation arithmetic, we use approximations to transcendental functions where applicable.
For example the arctangent for orientation angle computation uses a variation of a well-known quadratic approximation~\cite{rajan06}.

With the GPU implementation and inherent parallelism of our filters, we have observed order of magnitude speed-up over straightforward CPU implementation.
We observe linear performance scaling with the number of processed pixels for all implemented filtering operations and real-time performance for Full HD images on a mobile device (Fig.~\ref{fig:performance_both_per_filter} (bottom)) and up to 40MP images on a desktop PC (Fig.~\ref{fig:performance_both_per_filter} (top)).

\section{Conclusions}\label{sec:conclusions}

In this work, we presented an interactive framework for designing filter-based stylization which allows a designer to tune, modify, and conveniently explore the space of filters directly, empowering user creativity. Our framework is flexible, and not limited to any particular visual style.

In parallel to this manual design, we presented a procedural, fully automatic style creation process that follows a set of simple rules to generate hundreds of different styles. These styles can be selected through manual visualization or evaluated using a previously trained aesthetics quality assessment method (e.g. \cite{talebi2018nima}).
Our filtering and stylization framework was used to design filters for a free, experimental application called \textit{Google Storyboard}\footnote{\begin{tiny}\texttt{https://play.google.com/store/apps/details?id=com.google.android.apps.photolab.storyboard}\end{tiny}} that allows to automatically turn any video into a comic strip.

In order to make the system real-time, we apply the Best Linear Adaptive Enhancement (BLADE) framework for simple, trainable, and edge-adaptive filtering to realize fast approximations of sophisticated style effects. BLADE's computationally efficient inference allows for fully real-time applications on a mobile device and is easy to train and interpret.

For future work, we find several potential directions of explorations. First, one can include any new, more advanced filters. These may include CNN-based stylization approaches that would enrich the expressiveness of our system. A second possibility is to explore more complex hand-crafted features for filter selection. While this might increase the computational requirements, it would allow for more sophisticated and more content-adaptive stylization filters. Finally, the adaptation of the proposed approach to longer videos including episodic-length television and motion pictures, while enabling automatic or convenient change of style as required by the scene or the director, would be quite interesting. 

\bibliographystyle{iet}
\bibliography{stylization_2019}

\newcommand{\beginsupplement}{%
        \setcounter{table}{0}
        \renewcommand{\thetable}{S\arabic{table}}%
        \setcounter{figure}{0}
        \renewcommand{\thefigure}{S\arabic{figure}}%
     }

\beginsupplement
\clearpage
\appendix
\section*{Appendix}

\vspace{-0.2cm}
\begin{figure}[htb]
\centering
\includegraphics[width=0.75\linewidth]{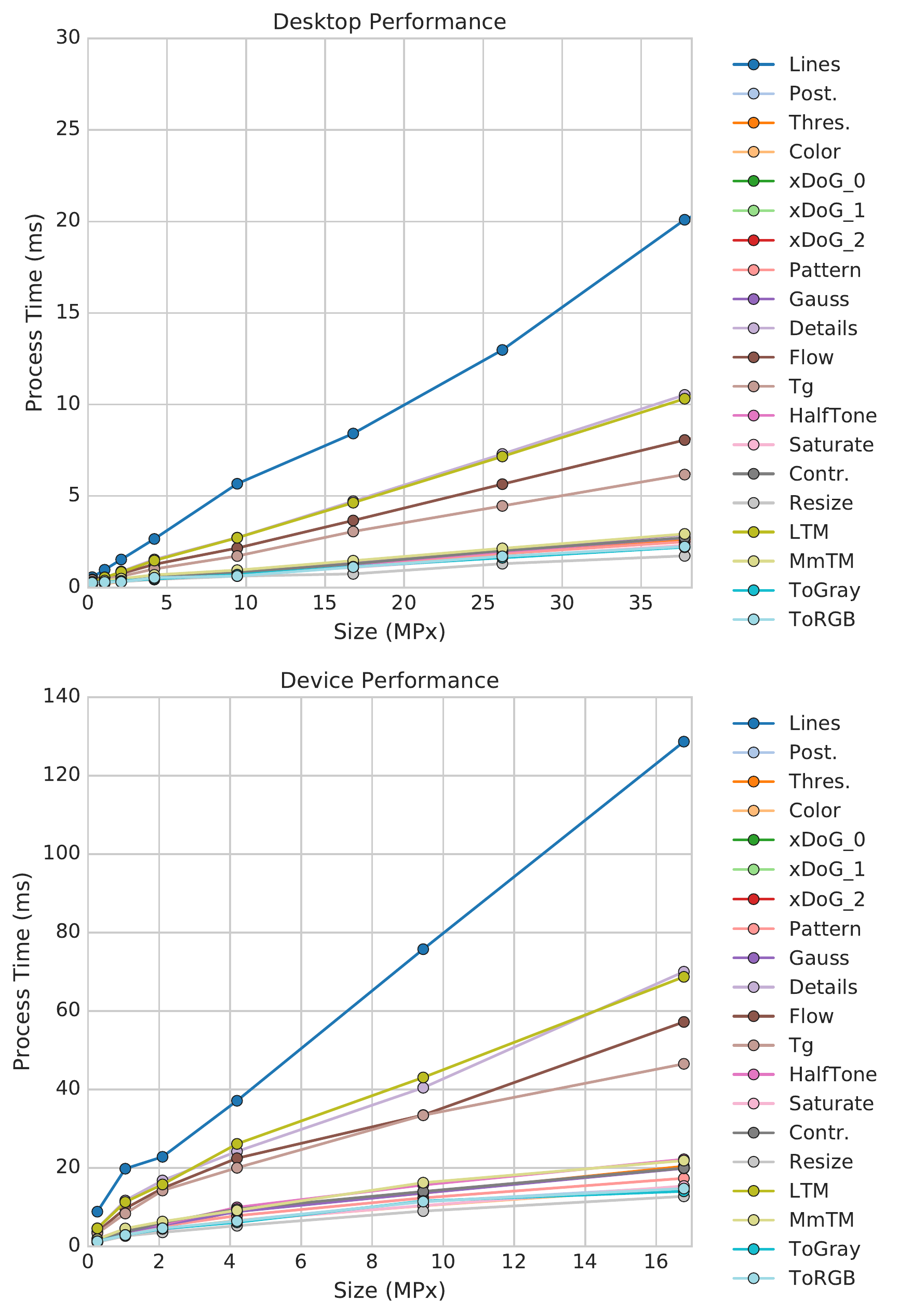}
\caption{\label{fig:performance_both_per_filter}Process time on GTX 1080 desktop PC and on \textit{`Device'} (Pixel 2018) vs. image size in megapixels for each of the filters in our system.}
\end{figure}

\vspace{-0.5cm}
\begin{figure}[htb]
\centering
\includegraphics[width=0.70\linewidth]{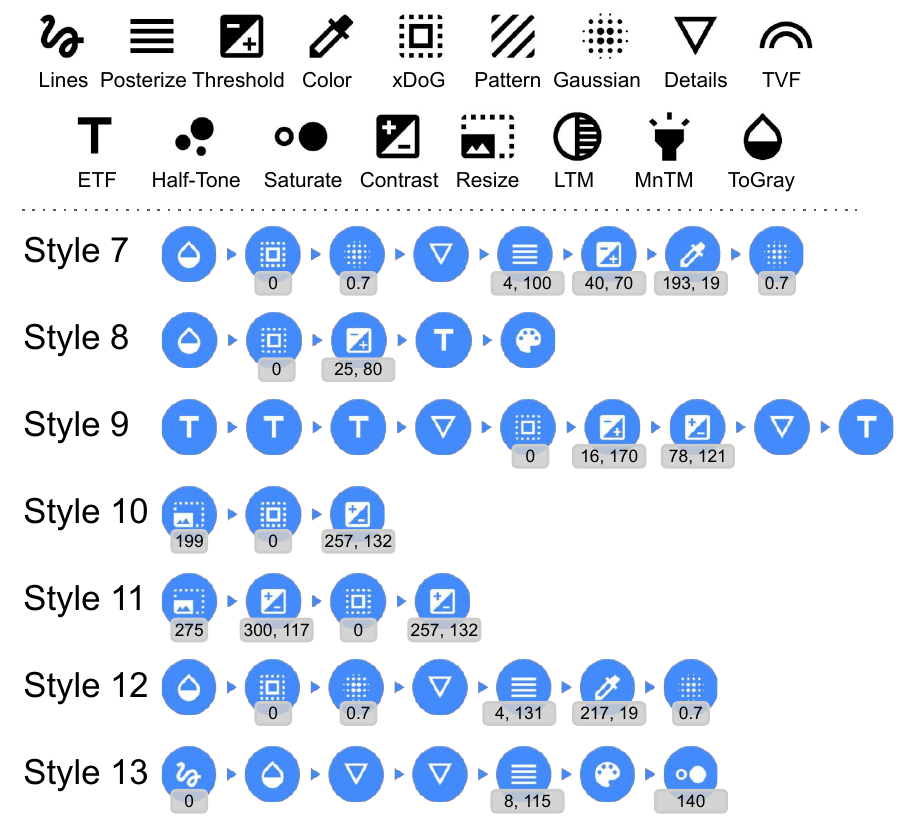}
\caption{\label{fig:2019_filters_2}Graph of each stylization. The caption of each filter are the parameters for the given filter.}
\end{figure}

\vspace{-0.5cm}
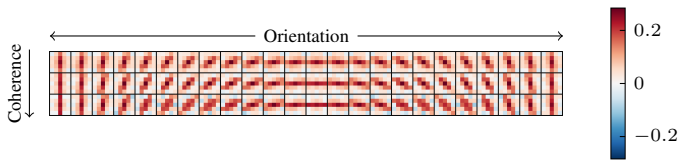
\begin{figure}[htb]
\centering
\mbox{%
\beginpgfgraphicnamed{images/etf_filters}%
\input{images/etf_filters.tikz}%
\endpgfgraphicnamed}
\caption{\label{fig:etf_filters} $5\times 5$ BLADE filters approximating edge
tangent flow with 24 different orientations and 3 coherence values.}
\end{figure}

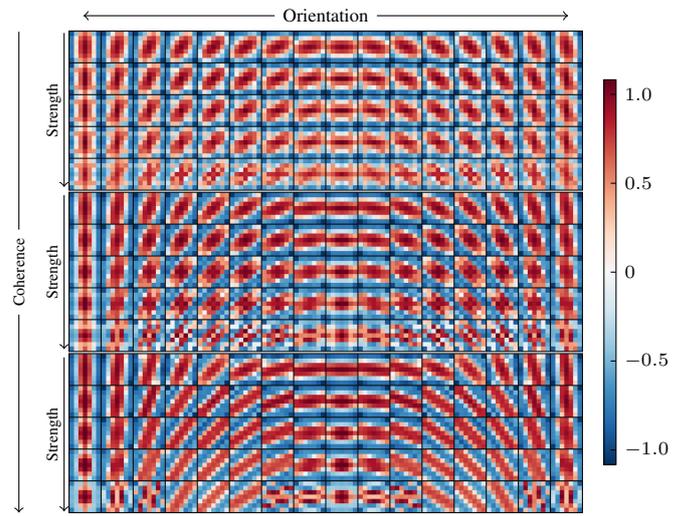
\begin{figure}[htb]
\centering
\mbox{%
\beginpgfgraphicnamed{images/fdog_filters}%
\input{images/fdog_filters.tikz}%
\endpgfgraphicnamed}
\caption{\label{fig:fdog_filters} $7\times 7$ BLADE filters for approximating Flow-XDoG with 16 different orientations, 5 strength bins, and 3 coherence bins.}
\end{figure}

\begin{figure}[htb]
\centering
\mbox{%
\beginpgfgraphicnamed{images/tv_flow_filters}%
\input{images/tv_flow_filters.tikz}%
\endpgfgraphicnamed}
\caption{\label{fig:tv_flow_filters} $7\times 7$ BLADE filters approximating
TV flow with 16 different orientations, 4 strength values, and 4 coherence
values.}
\end{figure}
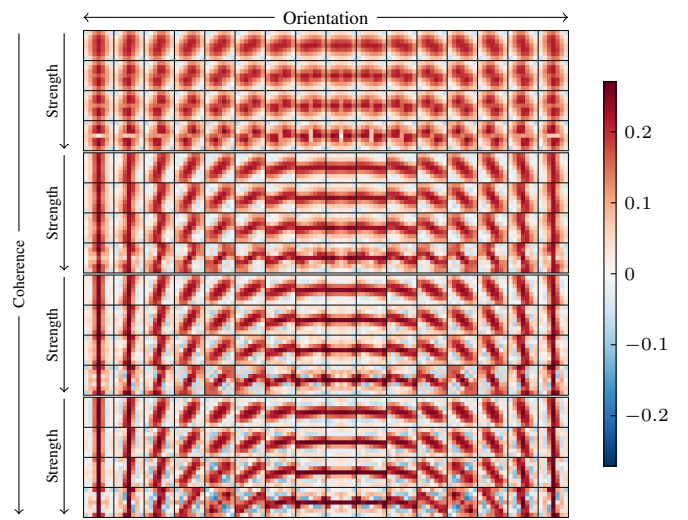

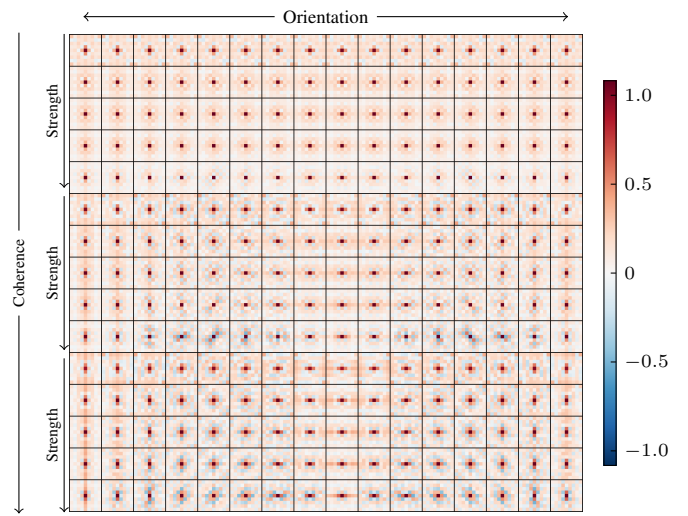
\begin{figure}[htb]
\centering
\mbox{%
\beginpgfgraphicnamed{images/details_-20_filters}%
\input{images/details_-20_filters.tikz}%
\endpgfgraphicnamed}
\caption{\label{fig:details_-20_filters} $9\times 9$ BLADE filters for approximating Detail Control -20 with 16 different orientations, 5 strength bins, and 3 coherence bins.}
\end{figure}

\begin{figure*}[tb]
\centering
\includegraphics[width=.95\linewidth]{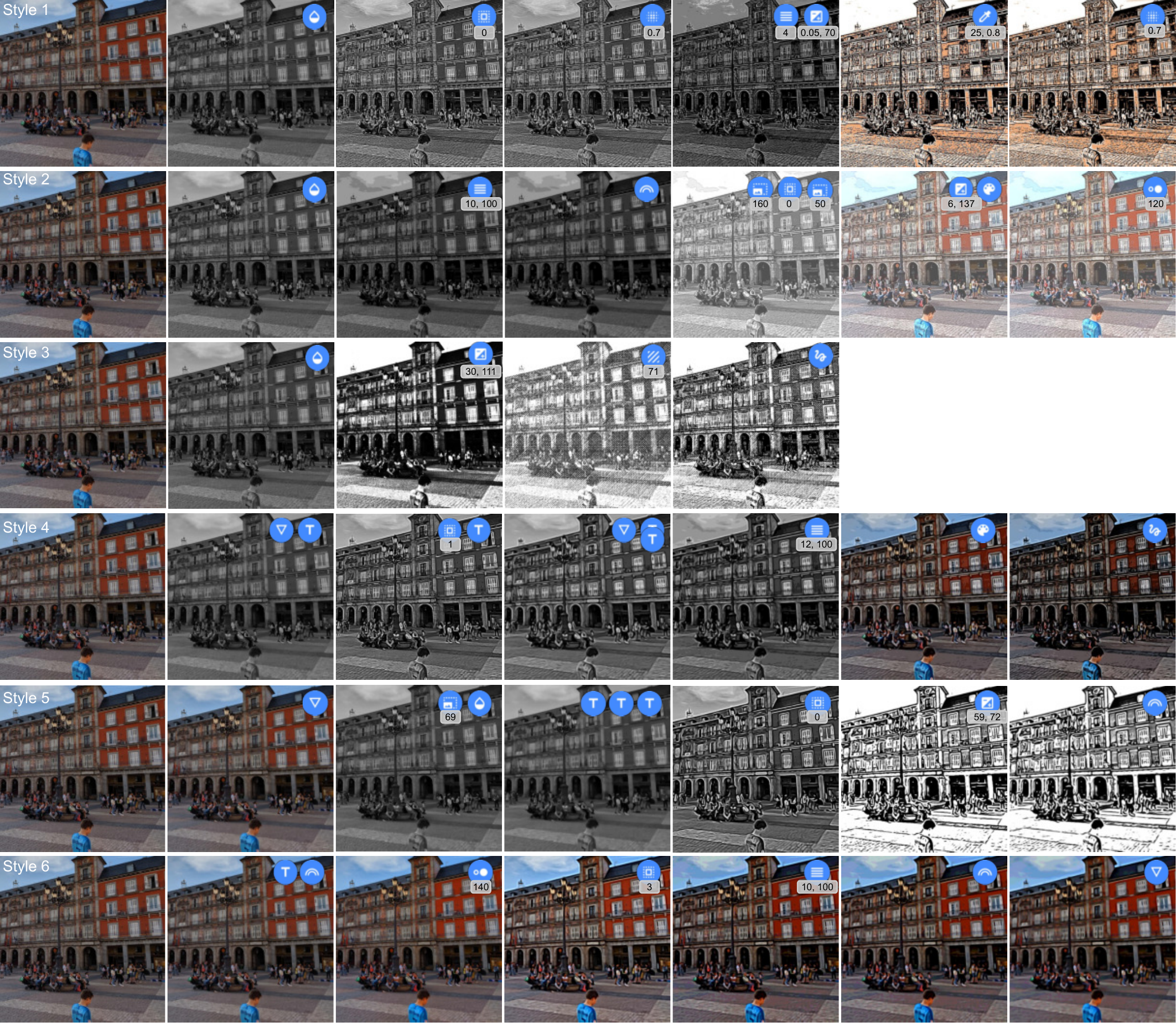}
\caption{\label{fig:2019_progress}Progression of each stylization from input (left) to the final stylization (right). The top right corner shows the filters applied in that step (see meaning of icons on Fig.~\ref{fig:2019_filters_2}). Note that \textit{Style 3} has just four filters. Readers are encouraged to zoom in aggressively (300\% or more).
}
\end{figure*}

\begin{figure*}[p]
\centering
\includegraphics[width=1.0\linewidth]{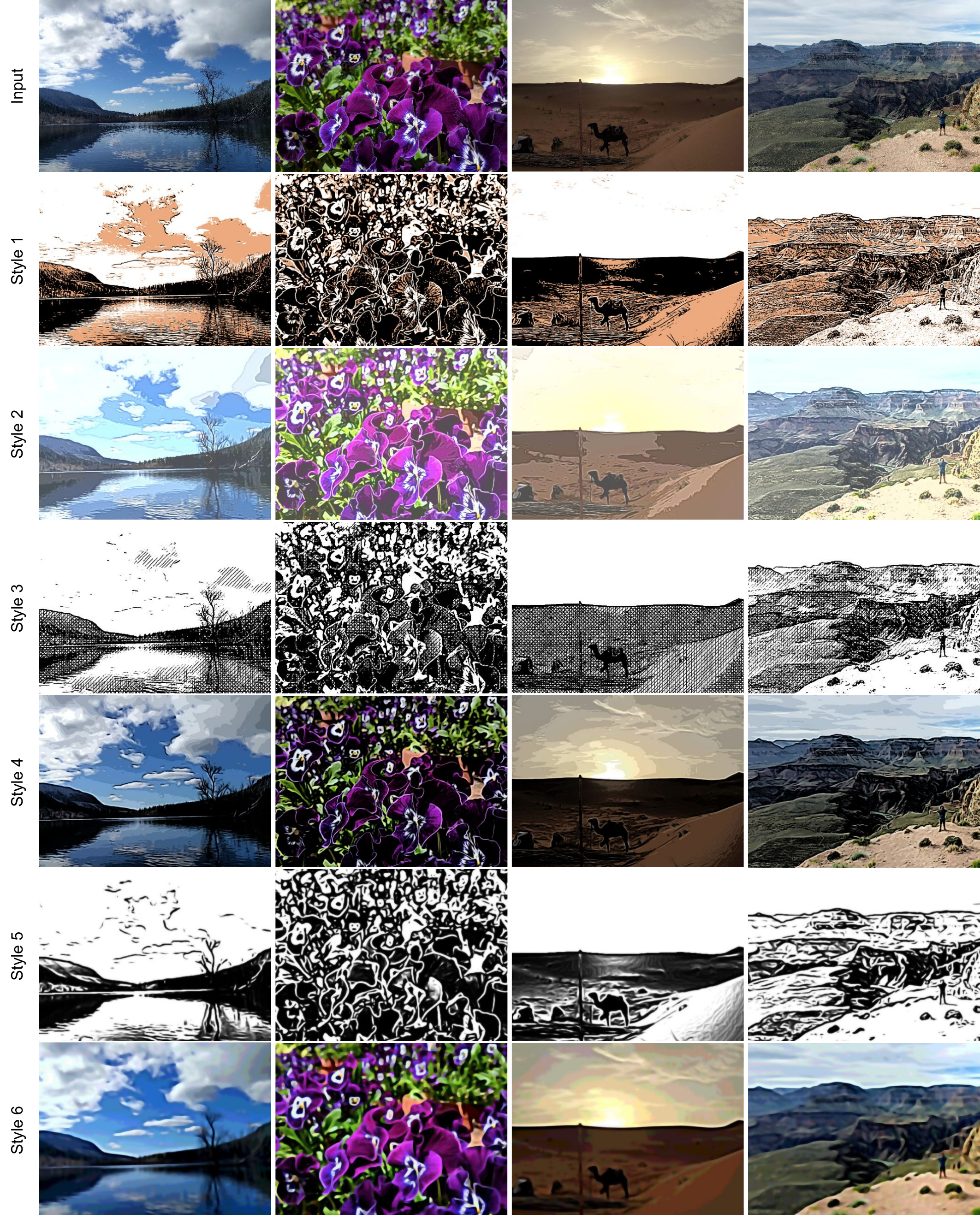}
\caption{\label{fig:2019_main_2}Additional stylization examples: Input image on top and six different stylizations created with our tool. See details in Sec.~\ref{subsec:new_stylizations}. Readers are encouraged to zoom in aggressively (200\% or more).}
\end{figure*}

\end{document}

%% file: images/inference_diagram.tikz

\begin{tikzpicture}[rotate=-90,scale=0.4,yscale=0.8,>=stealth',semithick]

\begin{scope}
\node [rectangle] (input) at (0,-3.75) {{\small Input patch}
$\vv{R}_i \vv{\inimage}$};
\coordinate (split0) at (0,0.4) {};

\begin{scope}[yshift=7.2cm]
\node [rectangle,draw,fill=PlotColorD,rounded corners=4pt,inner sep=4pt] (h1) at (-3.25,0.35)
{\small $\vv{\filter}^1$};
\node [rectangle,draw,fill=PlotColorD,rounded corners=4pt,inner sep=4pt] (h2) at (-0.85,0.35)
{\small $\vv{\filter}^2$};
\node (h3) at (1.25,0.35) {\small $\vdots$};
\node [rectangle,draw,fill=PlotColorD,rounded corners=4pt,inner sep=4pt] (h4) at (3.25,0.35)
{\small $\vv{\filter}^K$};
\end{scope}

\node [rectangle,draw,fill=PlotColorD,rounded corners=4pt,inner sep=3pt] (stanalysis) at (2,2.4)
{\small $s(i)$};

\draw (5,5.1) node [right] {\small Linear filterbank};
\draw (3.8,-1.5) node [right] {\small Filter selection};

\node [rectangle] (output) at (0,14) {{\small Output pixel} $\outimage_i$};

\begin{scope}[yshift=8.5cm]
\coordinate (split) at (0,1) {};
\draw [rounded corners=2.5pt] (split) -- (split-|h1) to (h1);
\draw (split-|h2) to (h2);
\draw (split) -- +(0.6,0);
\draw [dashed] (0.6,1) -- (1.6,1);
\draw [rounded corners=2.5pt] (1.6,1) -- (split-|h4) -- (h4);
\draw [->] (split) -- (output);
\end{scope}

\draw [rounded corners=2.5pt,->] (split0) -- (split0-|stanalysis) -- (stanalysis);

\begin{scope}[yshift=6cm]
\node [circle,draw,inner sep=1pt] (h1c) at (-3.35,0) {};
\node [circle,draw,inner sep=1pt] (h2c) at (-0.85,0) {};
\node [circle,draw,inner sep=1pt] (h4c) at (3.35,0) {};
\end{scope}

\node [circle,draw,inner sep=1pt] (sel) at (0,4) {};

\draw (h1) to (h1c);
\draw (h2) to (h2c);
\draw (h4) to (h4c);

\draw [xshift=0cm,yshift=4cm,very thick,gray!50] (40:1.2) arc (40:140:1.2);
\draw [shorten >=4pt] (sel) -- (h1c);
\draw [rounded corners=2.5pt,->,shorten >=2pt] (stanalysis) -- (stanalysis|-sel) -- (sel);
\end{scope}

\draw (input) -- (sel);

\end{tikzpicture}

%% file: images/quantization.tikz

\begin{tikzpicture}[scale=1.1]

\begin{scope}
\fill [ColorOri00!40] ((0:0) -- (84.375:1) arc (84.375:95.625:1)
  -- (95.625:-1) arc (95.625:84.375:-1) -- cycle;
\fill [ColorOri01!40] ((0:0) -- (73.125:1) arc (73.125:84.375:1)
  -- (84.375:-1) arc (84.375:73.125:-1) -- cycle;
\fill [ColorOri02!40] ((0:0) -- (61.875:1) arc (61.875:73.125:1)
  -- (73.125:-1) arc (73.125:61.875:-1) -- cycle;
\fill [ColorOri03!40] ((0:0) -- (50.625:1) arc (50.625:61.875:1)
  -- (61.875:-1) arc (61.875:50.625:-1) -- cycle;
\fill [ColorOri04!40] ((0:0) -- (39.375:1) arc (39.375:50.625:1)
  -- (50.625:-1) arc (50.625:39.375:-1) -- cycle;
\fill [ColorOri05!40] ((0:0) -- (28.125:1) arc (28.125:39.375:1)
  -- (39.375:-1) arc (39.375:28.125:-1) -- cycle;
\fill [ColorOri06!40] ((0:0) -- (16.875:1) arc (16.875:28.125:1)
  -- (28.125:-1) arc (28.125:16.875:-1) -- cycle;
\fill [ColorOri07!40] ((0:0) -- (5.625:1) arc (5.625:16.875:1)
  -- (16.875:-1) arc (16.875:5.625:-1) -- cycle;
\fill [ColorOri08!40] ((0:0) -- (-5.625:1) arc (-5.625:5.625:1)
  -- (5.625:-1) arc (5.625:-5.625:-1) -- cycle;
\fill [ColorOri09!40] ((0:0) -- (-16.875:1) arc (-16.875:-5.625:1)
  -- (-5.625:-1) arc (-5.625:-16.875:-1) -- cycle;
\fill [ColorOri10!40] ((0:0) -- (-28.125:1) arc (-28.125:-16.875:1)
  -- (-16.875:-1) arc (-16.875:-28.125:-1) -- cycle;
\fill [ColorOri11!40] ((0:0) -- (-39.375:1) arc (-39.375:-28.125:1)
  -- (-28.125:-1) arc (-28.125:-39.375:-1) -- cycle;
\fill [ColorOri12!40] ((0:0) -- (-50.625:1) arc (-50.625:-39.375:1)
  -- (-39.375:-1) arc (-39.375:-50.625:-1) -- cycle;
\fill [ColorOri13!40] ((0:0) -- (-61.875:1) arc (-61.875:-50.625:1)
  -- (-50.625:-1) arc (-50.625:-61.875:-1) -- cycle;
\fill [ColorOri14!40] ((0:0) -- (-73.125:1) arc (-73.125:-61.875:1)
  -- (-61.875:-1) arc (-61.875:-73.125:-1) -- cycle;
\fill [ColorOri15!40] ((0:0) -- (-84.375:1) arc (-84.375:-73.125:1)
  -- (-73.125:-1) arc (-73.125:-84.375:-1) -- cycle;

\foreach \th in {-84.375, -73.125, -61.875, -50.625, -39.375, -28.125, -16.875,
-5.625, 5.625, 16.875, 28.125, 39.375, 50.625, 61.875, 73.125, 84.375}
{
  \draw(\th:-1) -- (\th:-0.93) (\th:0.93) -- (\th:1);
}

\foreach \i/\th in {0/90, 1/78.75, 2/67.5, 3/56.25, 4/45, 5/33.75, 6/22.5,
7/11.25, 8/-0, 9/-11.25, 10/-22.5, 11/-33.75, 12/-45, 13/-56.25, 14/-67.5,
15/-78.75}
{
  \draw (\th:1.2) node {\tiny\sf \i};
}

\draw [semithick] (0,0) circle (1);

\draw (0,-1.6) node {\small\bf Orientation\vphantom{Sg}};
\end{scope}

\begin{scope}[xshift=2.3cm]

\begin{scope}[yshift=-0.85cm,xscale=1.5,yscale=1.7,xshift=0.0cm,semithick,xscale=-1,rotate=90]
\shade[top color=ColorStrengthTop,bottom color=ColorStrengthBottom,rounded corners=2.5pt] (0.0,-0.2) rectangle (1.0,0.1);

\fill[color=ColorStrengthTop!90!ColorStrengthBottom] (0.8,0.0) rectangle (1.0,0.2);
\fill[color=ColorStrengthTop!70!ColorStrengthBottom] (0.6,0.0) rectangle (0.8,0.2);
\fill[color=ColorStrengthTop!50!ColorStrengthBottom] (0.4,0.0) rectangle (0.6,0.2);
\fill[color=ColorStrengthTop!30!ColorStrengthBottom] (0.2,0.0) rectangle (0.4,0.2);
\fill[color=ColorStrengthTop!10!ColorStrengthBottom] (0.0,0.0) rectangle (0.2,0.2);

\draw (0.5,-0.15) node [left=2pt] {\tiny\sf In};
\draw (0.5,0.49) node [left,xshift=2pt] {\tiny\sf Out};

\draw (-0.1,0.2) node [left] {\footnotesize\sf 10};
\draw (1.1,0.2) node [left] {\footnotesize\sf 40};
\draw [semithick] (0,3pt) to (0,0) -- (1,0) to (1,3pt);

\foreach \x in {0.2, 0.4, 0.6, 0.8}
{
  \draw [thick] (\x,0) -- ++(0,3pt);
}

\end{scope}

\draw (0,-1.6) node {\small\bf Strength\vphantom{Sg}};
\end{scope}

\begin{scope}[xshift=4.3cm]

\begin{scope}[yshift=-0.85cm,xscale=1.5,yscale=1.7,xshift=0.0cm,semithick,xscale=-1,rotate=90]
\shade[top color=ColorCoherenceTop,bottom color=ColorCoherenceBottom,rounded corners=2.5pt] (0.0,-0.2) rectangle (1.0,0.1);

\fill[color=ColorCoherenceTop!83!ColorCoherenceBottom] (0.666,0.0) rectangle (1.0,0.2);
\fill[color=ColorCoherenceTop!50!ColorCoherenceBottom] (0.333,0.0) rectangle (0.666,0.2);
\fill[color=ColorCoherenceTop!17!ColorCoherenceBottom] (0.0,0.0) rectangle (0.333,0.2);

\draw [thick] (0,3pt) -- (0,0) node [left] {}
-- (1.0,0.0) node [left] {} -- ++(0,3pt);

\draw (-0.1,0.2) node [left] {\footnotesize\sf 0.2};
\draw (1.1,0.2) node [left] {\footnotesize\sf 0.8};

\foreach \x in {0.333, 0.667}
{
  \draw [semithick] (\x,0) -- ++(0,3pt);
}

\draw (0.5,-0.15) node [left=2pt] {\tiny\sf In};
\draw (0.5,0.49) node [left,xshift=2pt] {\tiny\sf Out};

\end{scope}

\draw (0,-1.6) node {\small\bf Coherence\vphantom{Sg}};
\end{scope}

\end{tikzpicture}

%% file: images/etf_filters.tikz

\begin{tikzpicture}

\begin{scope}[scale=0.9]
\node (0,0) {
\includegraphics[width=6.75cm]{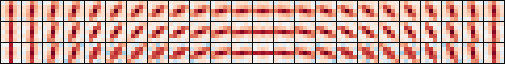}};

\node (ori) at (0,0.7) {\scriptsize Orientation};
\draw [->] (ori) -- +(-3.75,0);
\draw [->] (ori) -- +(3.75,0);

\draw [->] (-4.05,0.5) -- +(0,-1)
node [midway,xshift=0pt,above,rotate=90] {\scriptsize Coherence};
\end{scope}

\begin{scope}[xshift=4.1cm]
\node (0,0) {\includegraphics[width=0.16875cm,height=2cm]{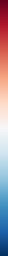}};
\draw (-0.09375,-1) rectangle (0.09375,1);

\foreach \y/\yl in {0/0,-0.70373/-0.2,0.70373/0.2}
{
  \begin{scope}[yshift=\y cm,xshift=0.09375cm]
  \pgftransformresetnontranslations
  \draw (-2pt,0) -- (0,0) node [right,black] {\scriptsize $\yl$};
  \end{scope}
}
\end{scope}

\end{tikzpicture}

%% file: images/fdog_filters.tikz

\begin{tikzpicture}[scale=0.85]

\node (0,0) {
\includegraphics[width=6.75cm]{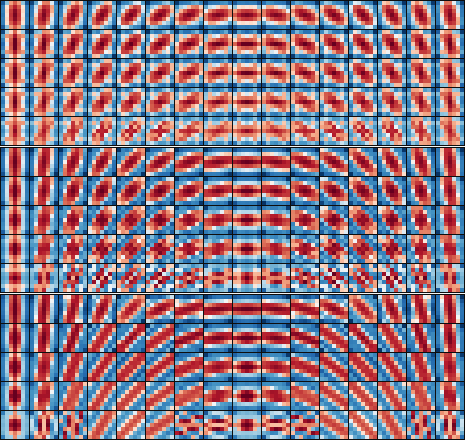}};

\node (ori) at (0,4) {\scriptsize Orientation};
\draw [->] (ori) -- +(-3.75,0);
\draw [->] (ori) -- +(3.75,0);

\foreach \y in {-1,0,1}
{
  \draw [->,yshift=\y * 2.53cm] (-4.05,1.2) -- +(0,-2.4)
  node [midway,xshift=2pt,above,rotate=90] {\tiny Strength};
}

\node (coh) at (-4.75,0) [rotate=90] {\tiny Coherence};
\draw (coh) -- +(0,3.75);
\draw [->] (coh) -- +(0,-3.75);

\begin{scope}[xshift=4.4cm]
\node (0,0) {\includegraphics[width=0.159375cm,height=5.1cm]{images/blue_red_colorbar.png}};
\draw (-0.09375,-3) rectangle (0.09375,3);

\foreach \y/\yl in {0/0,-1.38659/-0.5,1.38659/0.5,-2.77319/-1.0,2.77319/1.0}
{
  \begin{scope}[yshift=\y cm,xshift=0.09375cm]
  \pgftransformresetnontranslations
  \draw (-2pt,0) -- (0,0) node [right,black] {\scriptsize $\yl$};
  \end{scope}
}
\end{scope}

\end{tikzpicture}

%% file: images/tv_flow_filters.tikz

\begin{tikzpicture}[scale=0.85]

\node (0,0) {
\includegraphics[width=6.375cm]{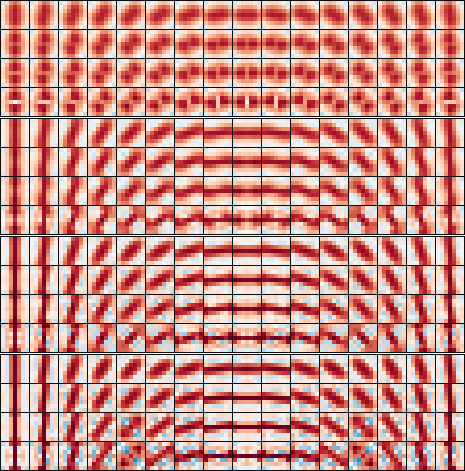}};

\node (ori) at (0,4) {\scriptsize Orientation};
\draw [->] (ori) -- +(-3.75,0);
\draw [->] (ori) -- +(3.75,0);

\foreach \y in {-1.5,-0.5,0.5,1.5}
{
  \draw [->,yshift=\y * 1.9cm] (-4.05,0.9) -- +(0,-1.8)
  node [midway,xshift=2pt,above,rotate=90] {\tiny Strength};
}

\node (coh) at (-4.75,0) [rotate=90] {\tiny Coherence};
\draw (coh) -- +(0,3.75);
\draw [->] (coh) -- +(0,-3.75);

\begin{scope}[xshift=4.4cm]
\node (0,0) {\includegraphics[width=0.159375cm,height=5.1cm]{images/blue_red_colorbar.png}};
\draw (-0.09375,-3) rectangle (0.09375,3);

\foreach \y/\yl in {0/0,-1.10701/-0.1,1.10701/0.1,-2.21402/-0.2,2.21402/0.2}
{
  \begin{scope}[yshift=\y cm,xshift=0.09375cm]
  \pgftransformresetnontranslations
  \draw (-2pt,0) -- (0,0) node [right,black] {\scriptsize $\yl$};
  \end{scope}
}
\end{scope}

\end{tikzpicture}

%% file: images/details_-20_filters.tikz

\begin{tikzpicture}[scale=0.85]

\node (0,0) {
\includegraphics[width=6.75cm]{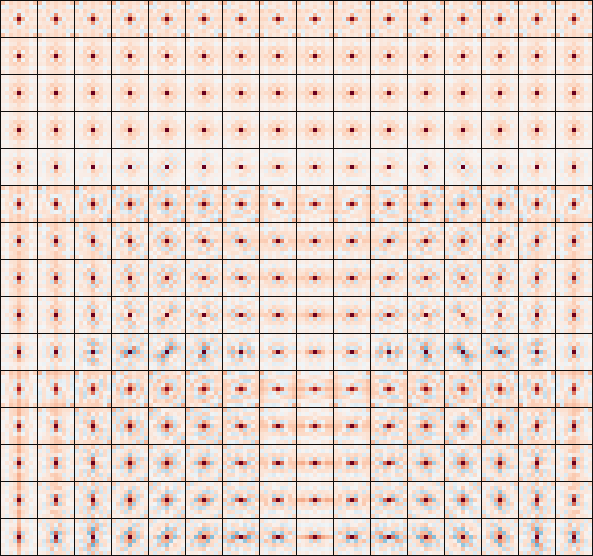}};

\node (ori) at (0,4) {\scriptsize Orientation};
\draw [->] (ori) -- +(-3.75,0);
\draw [->] (ori) -- +(3.75,0);

\foreach \y in {-1,0,1}
{
  \draw [->,yshift=\y * 2.53cm] (-4.05,1.2) -- +(0,-2.4)
  node [midway,xshift=2pt,above,rotate=90] {\tiny Strength};
}

\node (coh) at (-4.75,0) [rotate=90] {\tiny Coherence};
\draw (coh) -- +(0,3.75);
\draw [->] (coh) -- +(0,-3.75);

\begin{scope}[xshift=4.4cm]
\node (0,0) {\includegraphics[width=0.159375cm,height=5.1cm]{images/blue_red_colorbar.png}};
\draw (-0.09375,-3) rectangle (0.09375,3);

\foreach \y/\yl in {0/0,-1.38659/-0.5,1.38659/0.5,-2.77319/-1.0,2.77319/1.0}
{
  \begin{scope}[yshift=\y cm,xshift=0.09375cm]
  \pgftransformresetnontranslations
  \draw (-2pt,0) -- (0,0) node [right,black] {\scriptsize $\yl$};
  \end{scope}
}
\end{scope}

\end{tikzpicture}

%% file: main.bbl
\begin{thebibliography}{10}

\bibitem{movie_vincent}
Kobiela, D., Welchman, H.: `{Loving Vincent}', \emph{{Altitude Film
  Distribution}},  2017,

\bibitem{song_take}
Mansfield, T.: `Take on me', \emph{Warner Bros},  1984,

\bibitem{movie_dark}
Linklater, R.: `{A Scanner Darkly}', \emph{Warner Bros},  2006,

\bibitem{haeberli1990paint}
Haeberli, P.: ; ACM.
\newblock `Paint by numbers: Abstract image representations', \emph{ACM
  SIGGRAPH},  1990, \textbf{24}, (4), pp.~207--214

\bibitem{litwinowicz1997processing}
Litwinowicz, P.: ; Citeseer.
\newblock `Processing images and video for an impressionist effect',
  \emph{Computer Graphics and Interactive Techniques},  1997, pp.~ 407--414

\bibitem{kang2007coherent}
Kang, H., Lee, S., Chui, C.K.: ; ACM.
\newblock `Coherent line drawing', \emph{International symposium on
  Non-photorealistic animation and rendering},  2007, pp.~ 43--50

\bibitem{kyprianidis2008image}
Kyprianidis, J.E., D{\"o}llner, J.: `Image abstraction by structure adaptive
  filtering.', \emph{TPCG},  2008, pp.~ 51--58

\bibitem{kang2009flow}
Kang, H., Lee, S., Chui, C.K.: `Flow-based image abstraction', \emph{{IEEE
  Transactions on Visualization and Computer Graphics}},  2009, \textbf{15},
  (1), pp.~62--76

\bibitem{winnemoller2012xdog}
Winnem{\"o}ller, H., Kyprianidis, J.E., Olsen, S.C.: `{XDoG:} an {eXtended}
  difference-of-{Gaussians} compendium including advanced image stylization',
  \emph{Computers \& Graphics},  2012, \textbf{36}, (6), pp.~740--753

\bibitem{movie_dreams}
Ward, V.: `{What Dreams May Come}', \emph{Universal Studios},  1998,

\bibitem{movie_waking}
Linklater, R.: `{Waking Life}', \emph{Fox Pictures},  2001,

\bibitem{movie_sim_city}
Frank.Miller, R.R.: `{Sin City}', \emph{Miramax},  2005,

\bibitem{hu2014we}
Hu, Y., Manikonda, L., Kambhampati, S.: `What we instagram: A first analysis of
  instagram photo content and user types', \emph{{International AAAI conference
  on Web and Social Media}},  2014,

\bibitem{bakhshi2015we}
Bakhshi, S., Shamma, D.A., Kennedy, L., Gilbert, E.: `Why we filter our photos
  and how it impacts engagement', \emph{{International AAAI Conference on Web
  and Social Media}},  2015,

\bibitem{getreuer2018blade}
Getreuer, P., {Garcia-Dorado}, I., Isidoro, J., Choi, S., Ong, F., Milanfar,
  P.: ; IEEE.
\newblock `{BLADE:} filter learning for general purpose computational
  photography', \emph{ICCP},  2018, pp.~ 1--11

\bibitem{talebi2016fast}
Talebi, H., Milanfar, P.: `Fast multilayer {Laplacian} enhancement', \emph{IEEE
  Transactions on Computational Imaging},  2016, \textbf{2}, (4), pp.~496--509

\bibitem{milanfar2013tour}
Milanfar, P.: `A tour of modern image filtering new insights and methods, both
  practical and theoretical', \emph{IEEE Signal Processing Magazine},  2013,
  \textbf{30}, (1), pp.~106--128

\bibitem{milanfar2013symmetrizing}
Milanfar, P.: `Symmetrizing smoothing filters', \emph{SIAM, Journal on Imaging
  Science},  2013, \textbf{6}, (1), pp.~263--284

\bibitem{kyprianidis2013state}
Kyprianidis, J.E., Collomosse, J., Wang, T., Isenberg, T.: `A taxonomy of
  artistic stylization techniques for images and video', \emph{{IEEE
  Transactions on Visualization and Computer Graphics}},  2013, \textbf{19},
  (5), pp.~866--885

\bibitem{winnemoller2006real}
Winnem{\"o}ller, H., Olsen, S.C., Gooch, B.: `Real-time video abstraction',
  \emph{ACM Transactions On Graphics (ToG)},  2006, \textbf{25}, (3),
  pp.~1221--1226

\bibitem{barnes2015patchtable}
Barnes, C., Zhang, F.L., Lou, L., Wu, X., Hu, S.M.: `Patchtable: Efficient
  patch queries for large datasets and applications', \emph{ACM Transactions on
  Graphics (ToG)},  2015, \textbf{34}, (4), pp.~97

\bibitem{gatys2016image}
Gatys, L.A., Ecker, A.S., Bethge, M.: `Image style transfer using convolutional
  neural networks', \emph{IEEE Conference on Computer Vision and Pattern
  Recognition (CVPR)},  2016, pp.~ 2414--2423

\bibitem{elad2017style}
Elad, M., Milanfar, P.: `Style transfer via texture synthesis', \emph{IEEE
  Transactions on Image Processing},  2017, \textbf{26}, (5), pp.~2338--2351

\bibitem{gatys2017controlling}
Gatys, L.A., Ecker, A.S., Bethge, M., Hertzmann, A., Shechtman, E.:
  `Controlling perceptual factors in neural style transfer', \emph{IEEE
  Conference on Computer Vision and Pattern Recognition (CVPR)},  2017, pp.~
  3985--3993

\bibitem{johnson2016perceptual}
Johnson, J., Alahi, A., Fei.Fei, L.: ; Springer.
\newblock `Perceptual losses for real-time style transfer and
  super-resolution', \emph{European Conference on Computer Vision},  2016, pp.~
  694--711

\bibitem{frigo2016split}
Frigo, O., Sabater, N., Delon, J., Hellier, P.: `Split and match: Example-based
  adaptive patch sampling for unsupervised style transfer', \emph{IEEE
  Conference on Computer Vision and Pattern Recognition (CVPR)},  2016, pp.~
  553--561

\bibitem{liu2018proximal}
Liu, R., Fan, X., Cheng, S., Wang, X., Luo, Z.: `Proximal alternating direction
  network: A globally converged deep unrolling framework', \emph{Thirty-Second
  AAAI Conference on Artificial Intelligence},  2018,

\bibitem{chen2017trainable}
Chen, Y., Pock, T.: `Trainable nonlinear reaction diffusion: A flexible
  framework for fast and effective image restoration', \emph{IEEE Transactions
  on Pattern Analysis and Machine Intelligence},  2017, \textbf{39}, (6),
  pp.~1256--1272

\bibitem{lefkimmiatis2018universal}
Lefkimmiatis, S.: `Universal denoising networks: a novel {CNN} architecture for
  image denoising', \emph{IEEE Conference on Computer Vision and Pattern
  Recognition (CVPR)},  2018, pp.~ 3204--3213

\bibitem{corbineau2019learned}
Corbineau, M.C., Bertocchi, C., Chouzenoux, E., Prato, M., Pesquet, J.C.:
  `Learned image deblurring by unfolding a proximal interior point algorithm',
  \emph{IEEE International Conference on Image Processing (ICIP)},  2019, pp.~
  4664--4668

\bibitem{li2019algorithm}
Li, Y., Tofighi, M., Monga, V., Eldar, Y.C.: `An algorithm unrolling approach
  to deep image deblurring', \emph{IEEE International Conference on Acoustics,
  Speech and Signal Processing (ICASSP)},  2019, pp.~ 7675--7679

\bibitem{romano2017raisr}
Romano, Y., Isidoro, J., Milanfar, P.: `{RAISR: Rapid and Accurate Image Super
  Resolution}', \emph{IEEE Transactions on Computational Imaging},  2017,
  \textbf{3}, (1), pp.~110--125

\bibitem{talebi2018nima}
Talebi, H., Milanfar, P.: `Nima: Neural image assessment', \emph{IEEE
  Transactions on Image Processing},  2018, \textbf{27}, (8), pp.~3998--4011

\bibitem{sobel1990isotropic}
Sobel, I.: `An isotropic 3$\times$ 3 image gradient operator', \emph{Machine
  vision for three-dimensional scenes},  1990, pp.~ 376--379

\bibitem{louchet2011total}
Louchet, C., Moisan, L.: `Total variation as a local filter', \emph{SIAM
  Journal on Imaging Sciences},  2011, \textbf{4}, (2), pp.~651--694

\bibitem{forstner1987fast}
F{\"o}rstner, W., G{\"u}lch, E.: `A fast operator for detection and precise
  location of distinct points, corners and centres of circular features',
  \emph{Proceedings of the Intercomission Conference on Fast Processing of
  Photogrammetric Data},  1987, pp.~ 281--305

\bibitem{bigun1987optimal}
Bigun, J., Granlund, G.H.: `Optimal orientation detection of linear symmetry',
  \emph{IEEE First International Conference on Computer Vision},  1987, pp.~
  433--438

\bibitem{knutsson1993normalized}
Knutsson, H., Westin, C.F.: `Normalized and differential convolution',
  \emph{IEEE Conference on Computer Vision and Pattern Recognition (CVPR)},
  1993, pp.~ 515--523

\bibitem{weickert1998anisotropic}
Weickert, J.: `Anisotropic diffusion in image processing', \emph{Teubner
  Stuttgart},  1998,

\bibitem{zhu2009no}
Zhu, X., Milanfar, P.: `A no-reference sharpness metric sensitive to blur and
  noise', \emph{IEEE International Workshop on Quality of Multimedia
  Experience},  2009, pp.~ 64--69

\bibitem{takeda2006ICIP}
Takeda, H., Farsiu, S., Milanfar, P.: `Robust kernel regression for restoration
  and reconstruction of images from sparse noisy data', \emph{IEEE
  International Conference on Image Processing (ICIP)},  2006,

\bibitem{marquina2000explicit}
Marquina, A., Osher, S.: `Explicit algorithms for a new time dependent model
  based on level set motion for nonlinear deblurring and noise removal',
  \emph{SIAM Journal on Scientific Computing},  2000, \textbf{22}, (2),
  pp.~387--405

\bibitem{marr1980theory}
Marr, D., Hildreth, E.: `Theory of edge detection', \emph{Proceedings of the
  Royal Society of London Series B Biological Sciences},  1980, \textbf{207},
  (1167), pp.~187--217

\bibitem{muller2006procedural}
M{\"u}ller, P., Wonka, P., Haegler, S., Ulmer, A., Van.Gool, L.: ; ACM.
\newblock `Procedural modeling of buildings', \emph{ACM Transactions On
  Graphics (ToG)},  2006, \textbf{25}, (3), pp.~614--623

\bibitem{nishida2016interactive}
Nishida, G., Garcia.Dorado, I., Aliaga, D.G., Benes, B., Bousseau, A.:
  `Interactive sketching of urban procedural models', \emph{ACM Transactions on
  Graphics (ToG)},  2016, \textbf{35}, (4), pp.~130

\bibitem{rajan06}
Rajan, S., Wang, S., Inkol, R., Joyal, A.: `Efficient approximations for the
  arctangent function', \emph{IEEE Signal Processing Magazine},  2006, pp.~
  108--111

\end{thebibliography}
